\documentclass[preprint2]{aastex}

\shorttitle{Gas characteristics in the disk of Cen~A}
\shortauthors{Parkin et al.}

\begin{document}


\title{The physical characteristics of the gas in the disk of Centaurus~A using the \emph{Herschel Space Observatory}\footnote{\emph{Herschel} is an ESA space observatory with science instruments provided by European-led Principal Investigator consortia and with important participation from NASA.}}


\author{T.~J. Parkin\altaffilmark{2}, C.~D. Wilson\altaffilmark{2}, M.~R.~P. Schirm\altaffilmark{2}, M. Baes\altaffilmark{3}, M. Boquien,\altaffilmark{4}, A. Boselli\altaffilmark{4}, D. Cormier\altaffilmark{5}, M. Galametz\altaffilmark{6}, O.~\L. Karczewski\altaffilmark{7}, V. Lebouteiller\altaffilmark{8}, I. De~Looze\altaffilmark{3}, S.~C. Madden\altaffilmark{8}, H. Roussel\altaffilmark{9}, M.~W.~L. Smith\altaffilmark{10}, L. Spinoglio\altaffilmark{11}}
\altaffiltext{2}{Department of Physics \& Astronomy, McMaster University, Hamilton, Ontario, L8S~4M1, Canada; \email{parkintj@mcmaster.ca}}
\altaffiltext{3}{Sterrenkundig Observatorium, Universiteit Gent, Krijgslaan 281 S9, B-9000 Gent, Belgium}
\altaffiltext{4}{Laboratoire d'Astrophysique de Marseille, Universit\'e d'Aix-Marseille \& CNRS, UMR7326, 13388 Marseille Cedex 13, France}
\altaffiltext{5}{Institut f\"ur theoretische Astrophysik, Zentrum f\"ur Astronomie der Universit\"at Heidelberg, Albert-Ueberle Str. 2, D-69120 Heidelberg, Germany}
\altaffiltext{6}{Institute of Astronomy, University of Cambridge, Madingley Road, Cambridge, CB3 0HA, UK}
\altaffiltext{7}{Department of Physics \& Astronomy, University of Sussex, Brighton, BN1 9QH, UK}
\altaffiltext{8}{CEA, Laboratoire AIM, Universit\'e Paris VII, IRFU/Service d'Astrophysique, Bat. 709, Orme des Merisiers, 91191 Gif-sur-Yvette, France}
\altaffiltext{9}{Institut d'Astrophysique de Paris, UMR7095 CNRS, Universit\'e Pierre \& Marie Curie, 98 bis Boulevard Arago, F-75014 Paris, France}
\altaffiltext{10}{School of Physics \& Astronomy, Cardiff University, The Parade, Cardiff CF24 3AA, UK}
\altaffiltext{11}{Istituto di Astrofisica e Planetologia Spaziali, INAF-IAPS, Via Fosso  
del Cavaliere 100, I-00133 Roma, Italy}

\begin{abstract}
We search for variations in the disk of Centaurus~A of the emission from atomic fine structure lines using \emph{Herschel} PACS and SPIRE spectroscopy.  In particular we observe the [C~\textsc{ii}] (158~$\mu$m), [N~\textsc{ii}](122 and 205~$\mu$m), [O~\textsc{i}](63 and 145~$\mu$m) and [O~\textsc{iii}](88~$\mu$m) lines, which all play an important role in cooling the gas in photo-ionized and photodissociation regions.  We determine that the ([C~\textsc{ii}]+[O~\textsc{i}]$_{63}$)/$F_{\mathrm{TIR}}$ line ratio, a proxy for the heating efficiency of the gas, shows no significant radial trend across the observed region, in contrast to observations of other nearby galaxies.  We determine that 10--20\% of the observed [C\,\textsc{ii}] emission originates in ionized gas.  Comparison between our observations and a PDR model shows that the strength of the far-ultraviolet radiation field, $G_{0}$, varies between 10$^{1.75}$ and 10$^{2.75}$ and the hydrogen nucleus density varies between 10$^{2.75}$ and 10$^{3.75}$~cm$^{-3}$, with no significant radial trend in either property.   In the context of the emission line properties of the grand-design spiral galaxy M51 and the elliptical galaxy NGC~4125, the gas in Cen~A appears more characteristic of that in typical disk galaxies rather than elliptical galaxies.

\end{abstract}




\section{Introduction}
Nearby galaxies are excellent laboratories in which to study the properties of the cold interstellar medium (ISM), as the current capabilities of infrared and submillimeter observatories allow us to study them on sub-kiloparsec (kpc) scales.  In particular, we can investigate the origin of key far-infrared atomic fine-structure lines on these physical scales using the \emph{Herschel Space Observatory} \citep{2010A&A...518L...1P}.  Centaurus~A (Cen~A; NGC~5128), located only $3.8 \pm 0.1$~Mpc away \citep{2010PASA...27..457H}, is resolved at scales of a few hundred parsecs, thus giving us the opportunity to search for variations within the interstellar gas throughout the galaxy.

Cen~A (13$^{\mathrm{h}}$25$^{\mathrm{m}}$27.6$^{\mathrm{s}}$, $-43^{\circ}01\arcmin09\arcsec$) has an unusual morphology, as it is a giant elliptical that appears to have swallowed a smaller disk galaxy and estimates put this merger around 380~Myr ago \citep{1980ApJ...241..969T}.  The disk provides a prominent dust lane through the center, and shows a strong warp, giving it an `S' like shape at infrared wavelengths \citep{2002ApJ...565..131L, 2006ApJ...645.1092Q, 2008A&A...490...77W, 2012MNRAS.422.2291P}.  Cen~A is the closest galaxy with an active galactic nucleus (AGN) and associated radio jets extending approximately 4$^{\circ}$ in either direction \citep[e.g.][]{1997A&AS..121...11C, 1998A&ARv...8..237I}.  It is also rich in gas, both atomic (H~\textsc{i}) and molecular (H$_{2}$) hydrogen \citep{2008A&A...485L...5M,2010A&A...515A..67S}, as well as carbon monoxide (CO), as observed in various rotational transitions \citep{1987ApJ...322L..73P,1990ApJ...363..451E,1992ApJ...391..121Q,1993A&A...270L..13R, 2012MNRAS.422.2291P}.  For a detailed summary of the physical properties of the galaxy see \citet{1998A&ARv...8..237I} and \citet{2010PASA...27..463M}.

Recently, \citet{2012MNRAS.422.2291P} presented new photometric observations at 70, 160, 250, 350 and 500~$\mu$m using the Photodetector Array Camera and Spectrometer \citep[PACS;][]{2010A&A...518L...2P} and the Spectral and Photometric Imaging Receiver \citep[SPIRE;][]{2010A&A...518L...3G} on \emph{Herschel}.  Through dust spectral energy distribution (SED) modelling they found a radially decreasing trend in dust temperature from about 30 to 20~K.  Then they combined the resulting dust map with a gas map, created with CO($J=3-2$) observations from the James Clerk Maxwell Telescope (JCMT) and an H~\textsc{i} map \citep{2010A&A...515A..67S}, to produce a gas-to-dust mass ratio map.  This ratio also shows a radial trend from 275 near the center of the galaxy, decreasing to Galactic values of roughly 100 in the outer disk.  The high ratio in the center is attributed to local effects on the ISM from the AGN.  Here, we extend the investigation of the disk of Cen~A by combining the \emph{Herschel} PACS photometry with new PACS spectroscopic observations of important atomic fine structure lines to characterize the neutral and ionized gas.

Fine structure lines such as [C~\textsc{ii}](158~$\mu$m), [N~\textsc{ii}](122 and 205~$\mu$m), [O~\textsc{i}](63 and 145~$\mu$m) and [O\,\textsc{iii}](88~$\mu$m) (hereafter [C~\textsc{ii}], [N~\textsc{ii}]$_{122}$, [N~\textsc{ii}]$_{205}$, [O~\textsc{i}]$_{63}$, [O~\textsc{i}]$_{145}$ and [O\,\textsc{iii}], respectively) play a crucial role in the thermal balance of the gas in the ISM.  These lines provide a means of gas cooling by de-excitation via photon emission, rather than collisional de-excitation, which does not result in photon emission and thus inhibits gas cooling.  The [C~\textsc{ii}] line is a tracer of both neutral and ionized gas as C$^{+}$ is produced by far-ultraviolet (FUV) photons with energy greater than 11.26~eV, and it is one of the dominant coolants among the aforementioned lines with a luminosity of roughly 0.1--1~\% that of the far-infrared (FIR) luminosity in typical star-forming galaxies \citep[e.g.][]{1985ApJ...289..803S, 1993ApJ...404..219S, 2001ApJ...561..766M, 2011ApJ...728L...7G, parkin_2013}.  The two [O~\textsc{i}] lines trace neutral gas, while the [N~\textsc{ii}] and [O~\textsc{iii}] lines trace ionized gas.

A commonly used diagnostic of the heating efficiency of the gas is the
([C~\textsc{ii}]+[O~\textsc{i}]$_{63}$)/$F_{\mathrm{TIR}}$ (or sometimes [C~\textsc{ii}]/$F_{\mathrm{FIR}}$) line ratio, which represents the relative contributions of the FUV flux to the heating of gas versus dust, assuming [C~\textsc{ii}] and [O~\textsc{i}]$_{63}$ are the main coolants \citep{1985ApJ...291..722T}.  Observations show that as infrared color increases (thus increasing dust temperature), the heating efficiency decreases because the dust grains and polycyclic aromatic hydrocarbons (PAHs) that provide free electrons for gas heating via the photoelectric effect have become too positively charged to free electrons efficiently \citep{2001ApJ...561..766M, 2008ApJS..178..280B, 2011ApJ...728L...7G, 2012ApJ...747...81C, 2012A&A...544A..55B, 2012A&A...548A..91L, 2013A&A...549A.118C, parkin_2013}.

To determine the physical properties of the gas we need to compare ratios of our observed fine structure lines to those predicted by a PDR model.  There are a number of models which explore the characteristics of PDRs such as \citet{1986ApJS...62..109V,1988ApJ...334..771V}, \citet{1989ApJ...338..197S, 1995ApJS...99..565S}, \citet{1997ApJ...482..298L}, \citet{2000A&A...358..682S}, \citet{2006ApJS..164..506L} and \citet{2006A&A...451..917R}, but one of the most commonly used models was first developed by \citet{1985ApJ...291..722T}, consisting of a plane-parallel, semi-infinite slab PDR.  The gas is characterized by two free parameters, the hydrogen nucleus density, $n$, and the strength of the FUV radiation field, $G_0$, normalized to the Habing Field, $1.6 \times 10^{-3}$~erg~cm$^{-2}$~s$^{-1}$ \citep{1968BAN....19..421H}.  This model has now been updated by \citet{1990ApJ...358..116W}, \cite{1991ApJ...377..192H}, and \citet{1999ApJ...527..795K,2006ApJ...644..283K}.

Investigations of PDRs and cooling lines in Cen~A have previously been carried out by \citet{2000A&A...355..885U} and \citet{2001A&A...375..566N} using the Long Wavelength (LWS) spectrometer on the \emph{Infrared Space Observatory} (ISO).  \citet{2000A&A...355..885U} observed Cen~A at four pointings along the dust lane and found $G_{0} \sim 10^{2}$ and $n \sim 10^{3}$~cm$^{-3}$.  Using the same observations, \citet{2001A&A...375..566N} find $G_{0} = 10^{2.7}$ and $n \sim 10^{3.1}$~cm$^{-3}$.  In samples of normal star-forming galaxies, as well as samples including starburst, AGN, and star-forming galaxies such as those of \citet{2001ApJ...561..766M} and \citet{2001A&A...375..566N}, respectively, global values for $G_{0}$ range from 10$^{2}$ to 10$^{4.5}$ and $n$ ranges between 10$^{2}$ and 10$^{4.5}$~cm$^{-3}$.  In this paper, we look at the PDR characteristics of Cen~A on smaller scales (roughly 260~pc at the 14$\arcsec$ resolution of the JCMT) in search of radial variations.

The paper is organized as follows.  We describe our data processing for the spectroscopic observations in Section~\ref{Herschel_obs}, and discuss the general morphology of the various lines in Section~\ref{results}. In Section~\ref{pdrs} we compare our observations to theoretical models and discuss their implications. We compare the gas characteristics of Cen~A with M51 in Section~\ref{compare_m51} and summarize this work in Section~\ref{conclusions}.

\section{\emph{Herschel} Observations}\label{Herschel_obs}

\subsection{PACS spectroscopy}\label{pacs_spec}
\begin{figure}
\includegraphics[width=\columnwidth]{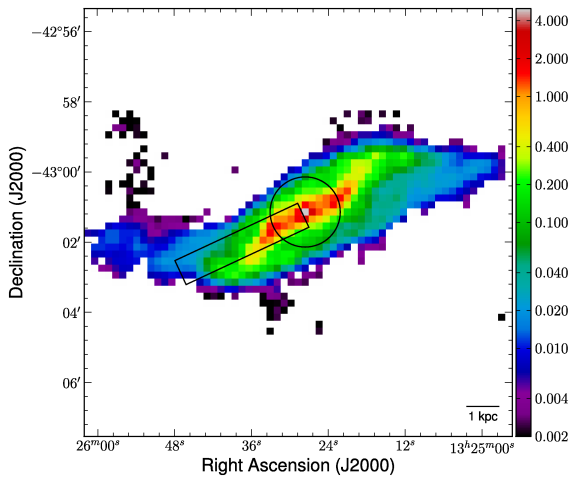}
\caption{The total infrared intensity calculated using Equation~\ref{eqn:Ftir} for Cen~A, at a resolution of 14$\arcsec$.  The \emph{Herschel} PACS footprint for our observations is shown as a rectangle while the circle outlines our SPIRE FTS footprint.  Units are in 10$^{-4}$~W~m$^{-2}$~sr$^{-1}$.}
\label{fig:F_tir}
\end{figure}

The data for the five fine structure lines observed with the PACS instrument were taken on 2011 July 9 using the unchopped grating scan mode.  They were taken as part of a \emph{Herschel} Guaranteed Time Key Project, the Very Nearby Galaxies Survey (VNGS; PI: C.~D. Wilson).  Each observation consists of a set of $7 \times 1$ footprints extending eastward along an orientation angle of 115$^{\circ}$ east of north.  One footprint covers a field-of-view of 47$\arcsec$ per side and the footprints are separated by 30$\arcsec$.  The PACS instrument consists of 25 spatial pixels (`spaxels') covering roughly 10$\arcsec$ on the sky each and thus we obtain 25 individual spectra per footprint.\footnote{PACS Observer's Manual (hereafter PACS~OM; HERSCHEL-HSC-DOC-0832, 2011), available from the ESA \emph{Herschel} Science Centre.}
The basic observational details for each line are summarized in Table~\ref{table: herschel_char}, while outlines of our observations are shown overlaid on a map of the total infrared intensity (see below for details) in Figure~\ref{fig:F_tir}.  We note that our observations do not cover the nucleus of Cen~A, as the nucleus was observed as part of another \emph{Herschel} Guaranteed Time project.

From Level~0 to Level~2 the PACS spectroscopic observations are processed with the standard pipeline for unchopped scans using the Herschel Interactive Processing Environment \citep[HIPE; ][]{2010ASPC..434..139O} version 9.2 with calibration files FM,41.  For details of the pipeline see \citet{parkin_2013} or the PACS Data Reduction Guide.\footnote{Available for download from the ESA \emph{Herschel} Science Centre. http://herschel.esac.esa.int/hcss-doc-10.0/index.jsp\#pacs\_spec:pacs\_spec}  Level~2 cubes are exported to PACSman v3.52 \citep{2012A&A...548A..91L} where each individual spectrum is fit with a second order polynomial and Gaussian function for the baseline and line, respectively.  Lastly, we create a map by projecting the rasters onto a common grid with a pixel scale of 3.133$\arcsec$.  In Figure~\ref{fig:pacs_spec_maps} we show the final mosaicked observations for the [C~\textsc{ii}], [N~\textsc{ii}]$_{122}$, [O~\textsc{i}]$_{63}$, [O~\textsc{i}]$_{145}$ and [O~\textsc{iii}] fine structure lines at their native resolution.

\begin{deluxetable}{lcccccc}
\tabletypesize{\small}
\tablecolumns{7}
\tablecaption{Details of the \emph{Herschel} spectroscopic observations of Centaurus~A\label{table: herschel_char}}
\tablewidth{0pt}
\tablehead{
\colhead{Line} & \colhead{Wavelength} & \colhead{OBSID} & \colhead{Date of} & \colhead{Map Size}
 	& \colhead{FWHM\tablenotemark{a}} & \colhead{Integration} \\
                   & \colhead{($\mu$m)}   &            & \colhead{Observation}
    & \colhead{$\arcmin\times\arcmin$} & \colhead{($\arcsec$)} & \colhead{Time (s)}}
 \startdata
 [O\,\textsc{i}]       & 63.184     & 1342223819 & 2011 Jul 9 & 0.72~$\times$~4.0
 	& $\sim$9.3             & 1886 \\
 $[$O\,\textsc{iii}$]$ & 88.356     & 1342223817 & 2011 Jul 9 & 0.72~$\times$~4.0
 	& $\sim$9.3             & 3194 \\
 $[$N\,\textsc{ii}$]$  & 121.898    & 1342223818 & 2011 Jul 9 & 0.72~$\times$~4.0
 	& $\sim$10              & 3198  \\
 $[$O\,\textsc{i}$]$   & 145.525    & 1342223815 & 2011 Jul 9 & 0.72~$\times$~4.0
 	& $\sim$11              & 5840  \\
 $[$C\,\textsc{ii}$]$  & 157.741    & 1342223816 & 2011 Jul 9 & 0.72~$\times$~4.0
 	& $\sim$11.5            & 1886  \\
 $[$N\,\textsc{ii}$]$  & 205.178    & 1342204036 & 2010 Aug 23 & $\sim2\arcmin$~diameter circle
 	& $17$                  & 17843 \\
 \enddata
 \tablenotetext{a}{Values are from the PACS Observer's Manual and the SPIRE Observers' Manual.}
\end{deluxetable}

\subsection{SPIRE spectroscopy}\label{spire_spec}
The SPIRE Fourier Transform Spectrometer (FTS) observation of Cen~A consists of a fully sampled map at high spectral resolution.  The [N\,\textsc {ii}]$_{205}$ line comes from observations using the SPIRE short wavelength (SSW) bolometer array, consisting of 37 hexagonally arranged bolometers with a combined total field-of-view of 2.6$\arcmin$ (although only bolometers within the central $\sim 2.0\arcmin$ are well calibrated).\footnote{Hereafter SPIRE~OM.  Document HERSCHEL-DOC-0798 version 2.4 (June 2011), is available from the ESA \emph{Herschel} Science Centre.}

We processed the observation using HIPE v11.0 developer's build 2652 and calibration set v10.1 using the standard pipeline (see \citet{parkin_2013} for details).  Next, we built a spectral cube using the HIPE function ``spireProjection()" with the naive projection option, then we fit the spectral line in each pixel of the cube with a sinc function.  Finally, a map is produced by integrating over each line at a resolution of $\sim 16\arcsec$ and with a 12$\arcsec$ pixel scale.  We chose this pixel size to match the common pixel size we adopted for all of the maps.  The final [N\,\textsc{ii}]$_{205}$ map at its native resolution is shown in Figure~\ref {fig:pacs_spec_maps}.  We note that in addition to the [N\,\textsc {ii}]$_{205}$ line, the FTS spectrum reveals detections in the CO ladder from the CO($J=4-3$) rotational transition at 461.04~GHz (650.25~$\mu$m) up to CO($J=8-7$) rotational transition at 921.80~GHz (325.23~$\mu$m), as well as the two [C\,\textsc{i}] lines at 492.16~GHz (609.14~$\mu$m) and 809.34~GHz (370.42~$\mu$m) (see Figure 3 in \citet{2014A&A...562A..96I}, which displays the full FTS spectrum from the central pixel).  However, a full FTS spectral analysis is beyond the scope of this paper and will be presented in a later paper.

\subsection{Ancillary Data}
We also use previously published PACS photometry at 70 and 160~$\mu$m \citep{2012MNRAS.422.2291P}.  These data have been reprocessed up to Level 1 using the PACS photometer pipeline \citep{2009ASPC..411..531W} in HIPE v9.0 (calibration file set FM,41), and then passed into \textsc{Scanamorphos} v21 \citep{2013PASP..125.1126R}, which was used to produce the final maps.  These maps were set to a final pixel scale of 1.4 and 2.85$\arcsec$ at 70 and 160~$\mu$m, respectively.  From the same paper we also make use of the CO($J=3-2$) observations taken at the JCMT.  We refer the reader to this paper for details on how the CO ($J=3-2$) map was constructed.  Lastly, we make use of the \emph{Spitzer} MIPS 24~$\mu$m data, reprocessed as described in \citet{2012MNRAS.423..197B}.

\subsection{Convolution and Re-sampling}\label{prep_analysis}
The PACS spectroscopic maps were convolved to a common resolution matching that of our CO($J = 3-2$) observations from the JCMT (14$\arcsec$) using Gaussian kernels.  The MIPS 24~$\mu$m and PACS 70 and 160~$\mu$m maps were convolved to the same resolution using the appropriate kernels from \citet{2011PASP..123.1218A}.  All of our maps were resampled onto a map with a pixel scale of 12$\arcsec$, such that each pixel is mostly independent.  Lastly, we mask out all detections below 5$\sigma$ in our spectroscopic maps to ensure robust line ratios for our analysis.

Calibration uncertainties are 4\% for the MIPS 24~$\mu$m photometry\footnote{MIPS Instrument Handbook, available at http://irsa.ipac.caltech.edu/data/SPITZER/docs/mips/\-mipsinstrumenthandbook/home/} and 5\% for the PACS 70 and 160~$\mu$m maps (PACS~OM).  The PACS spectroscopic maps have 30\% calibration uncertainties (PACS~OM) while the SPIRE FTS map has a 7\% calibration uncertainty (SPIRE~OM).  We note that calibration uncertainties are included in the reported errors unless otherwise stated.

\section{Map Analysis}\label{results}

\subsection{Morphological Properties of the Line Emission}\label{properties}
\begin{figure*}
 \includegraphics[width=7.2cm]{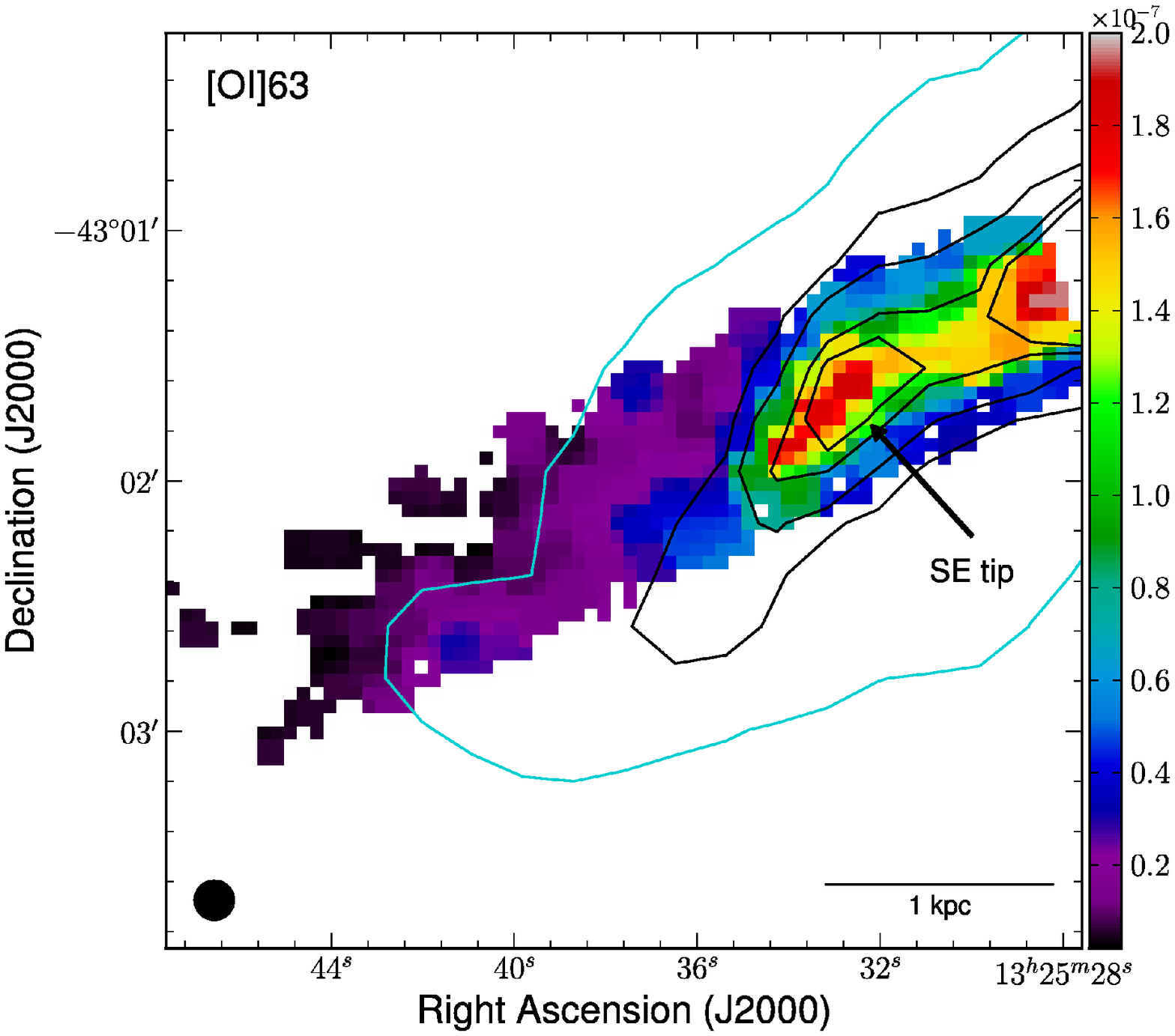}
 \includegraphics[width=7.2cm]{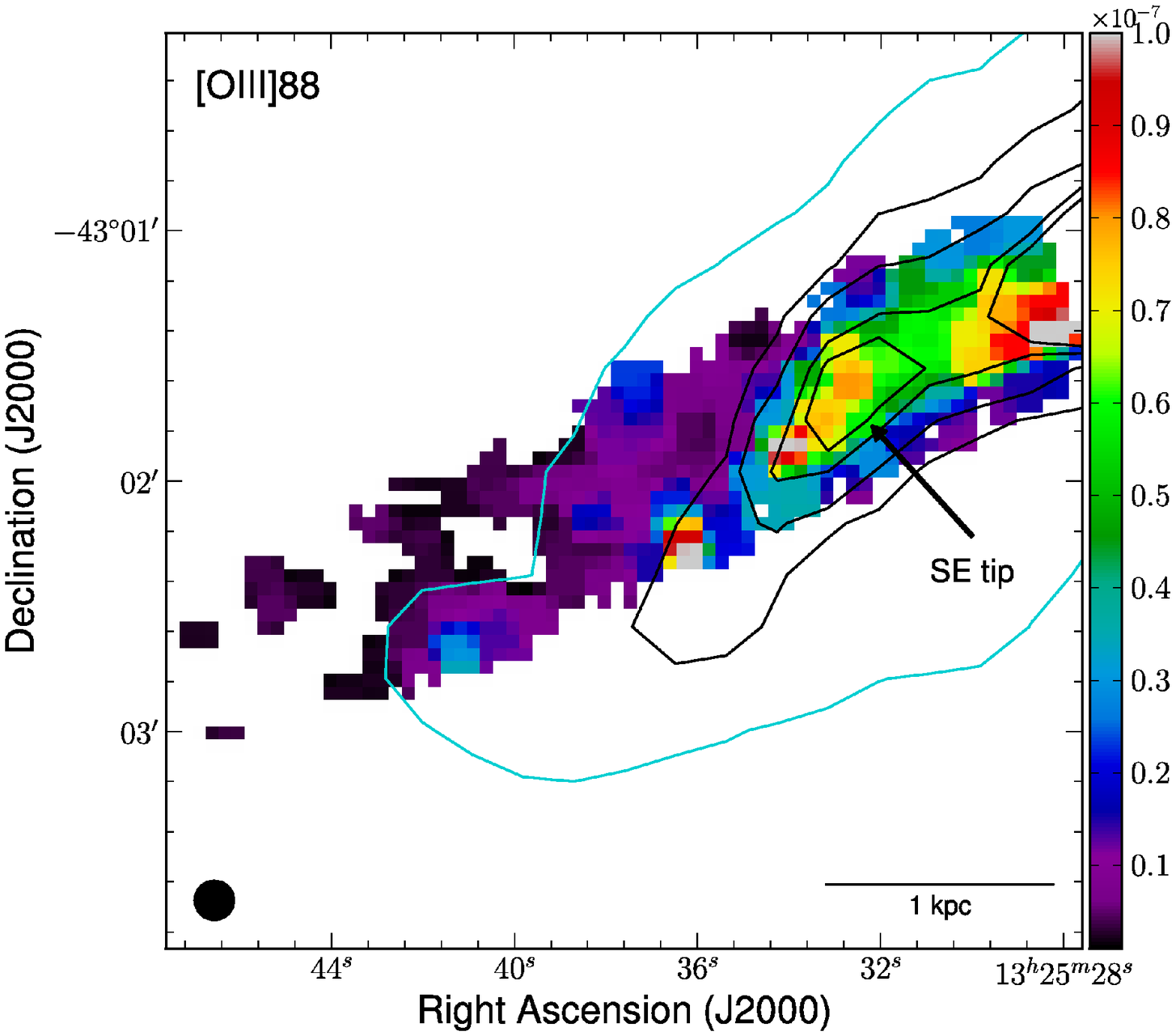}
 \includegraphics[width=7.2cm]{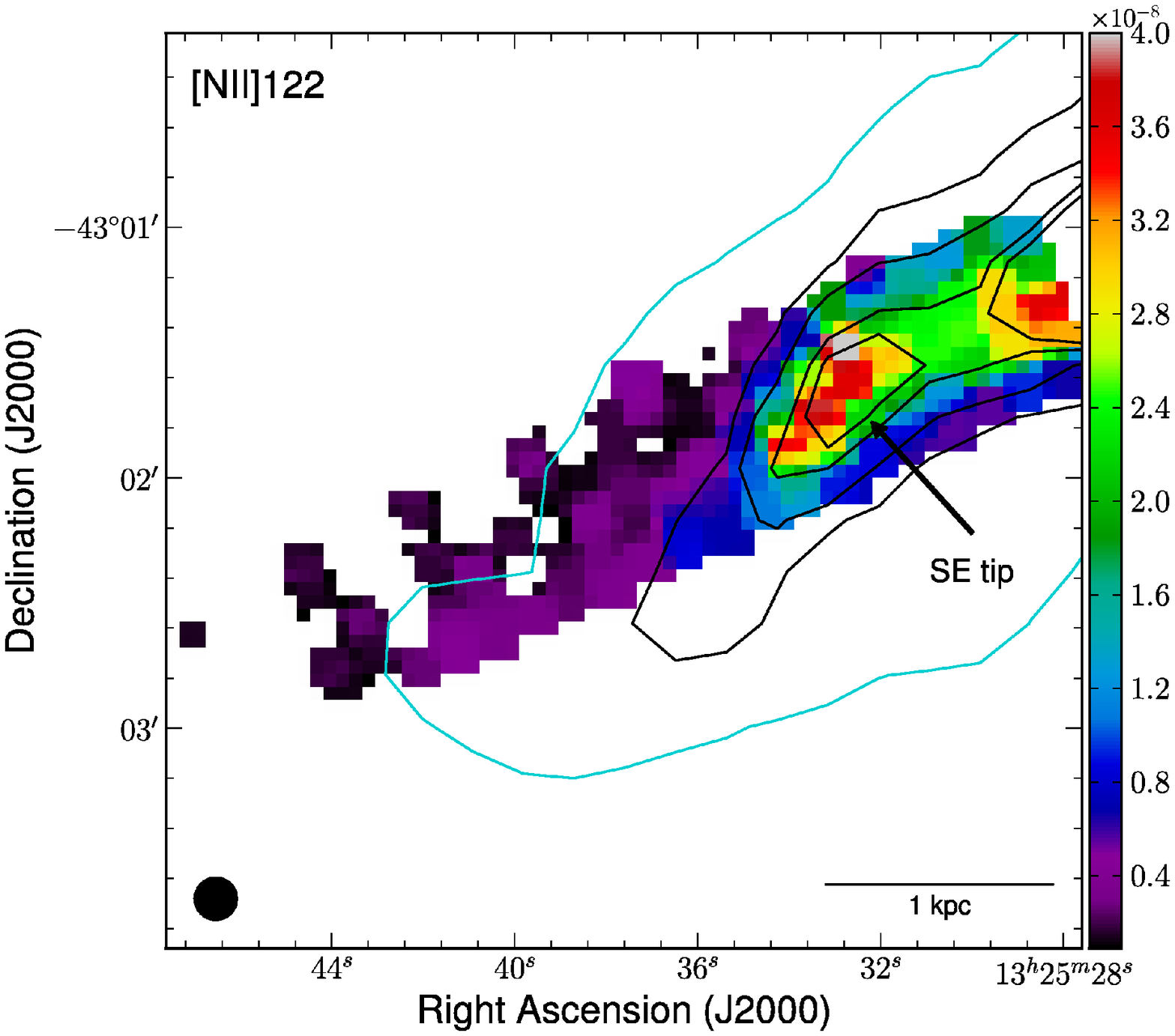}
 \includegraphics[width=7.2cm]{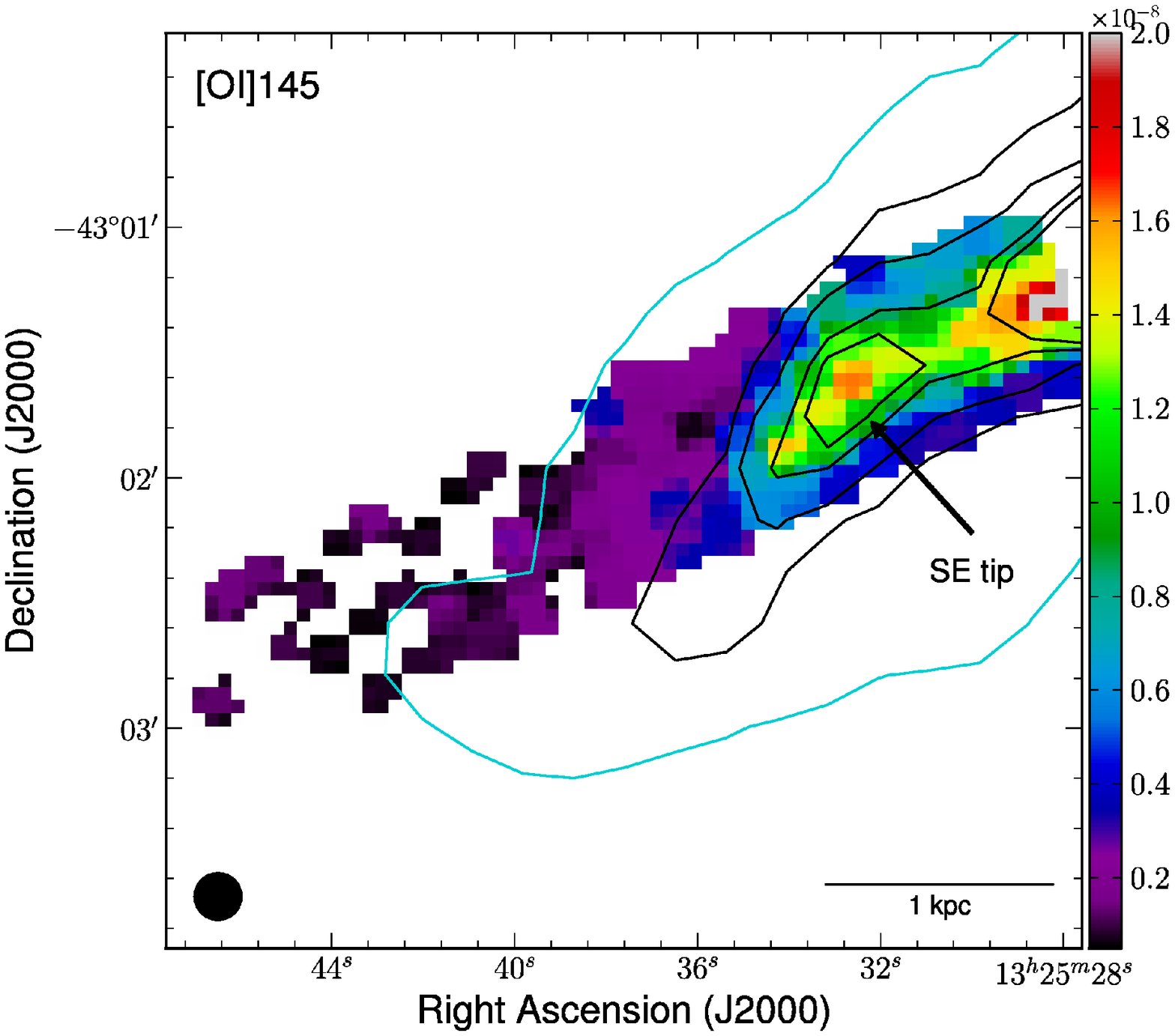}
 \includegraphics[width=7.2cm]{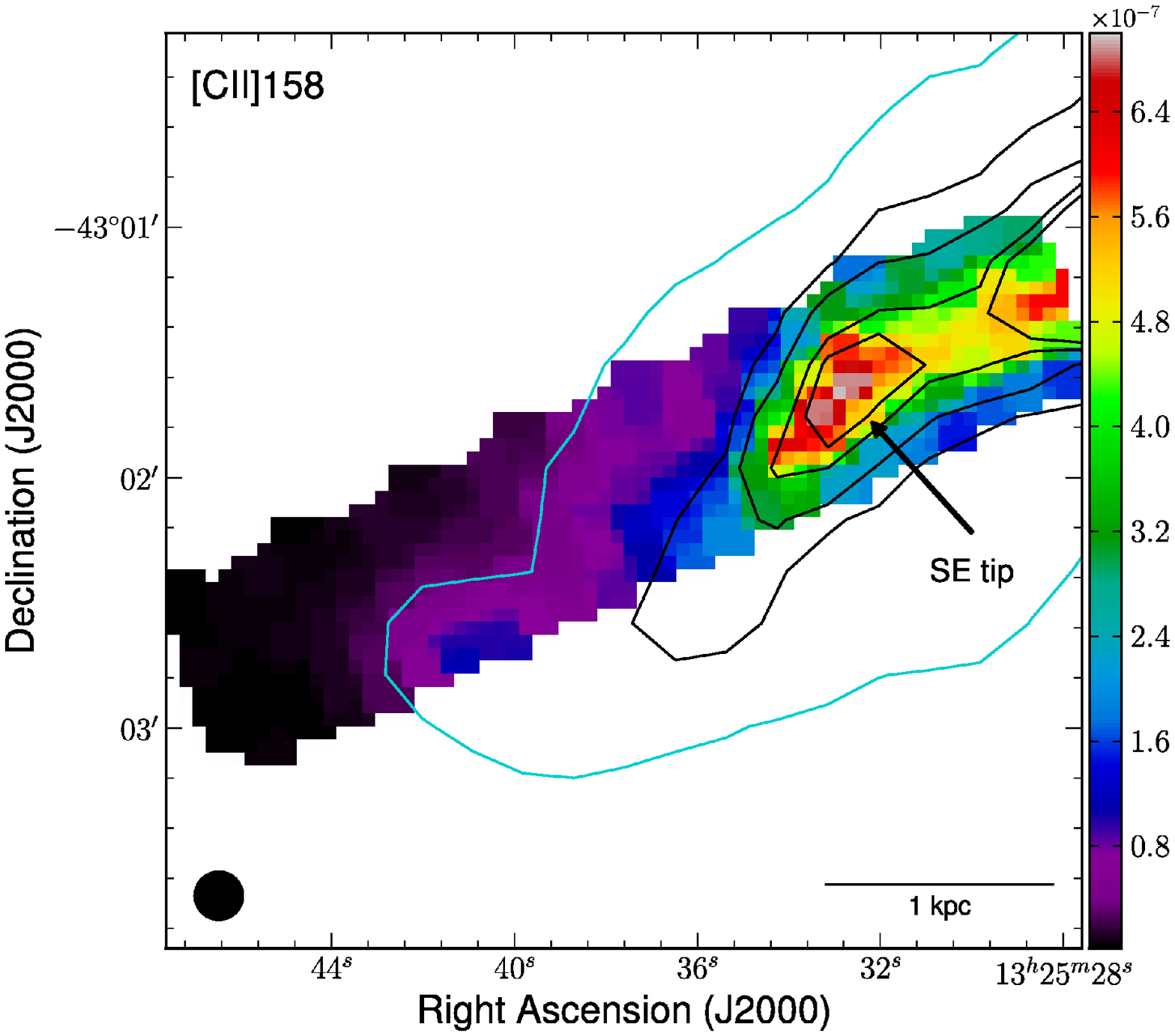}
 \includegraphics[width=7.2cm]{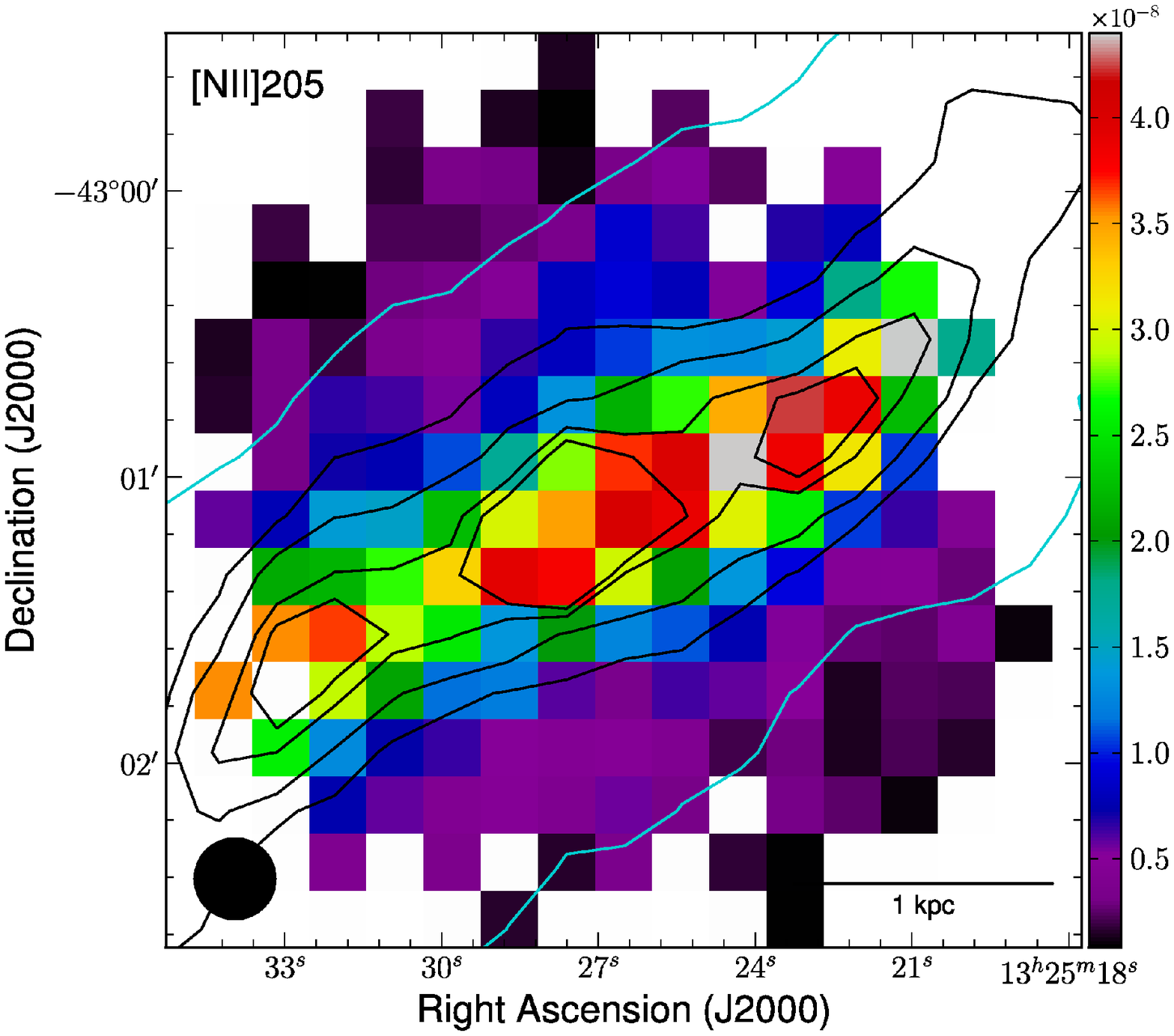}
\caption{The maps of our \emph{Herschel} PACS and SPIRE spectroscopic observations of the far-infrared cooling lines at their native resolution and pixel scale.  North is up and east is to the left in all panels.  We have applied a 3$\sigma$ cut to these maps to highlight robust detections; however, we note that in our analysis we use a 5$\sigma$ cut to ensure robust line ratios.  Units in all images are W~m$^{-2}$~sr$^{-1}$.  Contours from the \emph{Herschel} PACS 70~$\mu$m map tracing warm dust are overlaid on top with the levels corresponding to $3 \times 10^{-6}$, $1.5 \times 10^{-5}$, $3.0 \times 10^{-5}$, $6.0 \times 10^{-5}$, and $7.5 \times 10^{-5}$~W~m$^{-2}$~sr$^{-1}$.  The beam size for each line is shown as a black filled circle in the lower left corner.  The 'SE tip' referred to in the text is denoted by an arrow.}
\label{fig:pacs_spec_maps}
\end{figure*}

Figure~\ref{fig:pacs_spec_maps} shows our PACS and SPIRE spectroscopic maps at their native resolution and pixel scale.  We note that these maps are in units of integrated intensity and only have a 3$\sigma$ cut applied for display purposes; however, all flux measurements and analyses are carried out on the 5$\sigma$ cut maps.  The [C~\textsc{ii}] emission, tracing both neutral and ionized gas, shows a smooth decrease from near the center of the galaxy to the edge of our map.  Peaks in the [C~\textsc{ii}] emission correspond to peaks in the warm dust emission as traced by the 70~$\mu$m emission, overlaid as contours Figure~\ref {fig:pacs_spec_maps}.  The strongest emission is a factor of roughly 100 times higher than the outer part of the map and we denote this peak the  'SE tip' in Figure~\ref{fig:pacs_spec_maps}. The peak at the SE tip has also been seen previously in the \emph{Herschel} PACS 160~$\mu$m band as well as the three SPIRE photometric bands at 250, 350, and 500~$\mu$m, and in CO($J=3-2$) emission \citep{2012MNRAS.422.2291P}.  Furthermore, \citet{2006ApJ...645.1092Q} presented \emph{Spitzer} Infrared Array Camera (IRAC) photometry that demonstrates a parallelogram shaped ring, coincident with our [C~\textsc{ii}] observations.  The total flux in our [C~\textsc{ii}] map is $(4.27 \pm 1.28) \times 10^{-14}$~W~m$^{-2}$ over an area of approximately 11200~square arcseconds.  Our value is in fairly good agreement with \citet{2000A&A...355..885U}, who found a total flux for their center and south-east pointings (those which overlap our observations) of $3.83 \times 10^{-14}$~W~m$^{-2}$ covering a total area of 11100~square arcseconds given ISO's 70$\arcsec$ beam.  Any disagreement likely is due to the fact that our observations are not entirely co-spatial with theirs.

The [O~\textsc{i}]$_{63}$ and [O~\textsc{i}]$_{145}$ maps reveal the distribution of neutral gas, and as for [C~\textsc{ii}] we observe a radial decrease in intensity.  The total flux in these lines is $(1.17 \pm 0.35) \times 10^{-14}$~W~m$^{-2}$ and $(1.0 \pm 0.3) \times 10^{-15}$~W~m$^{-2}$ at 63 and 145~$\mu$m, respectively.  Interestingly, the  [O~\textsc{i}] emission peaks in the innermost region, unlike the [C~\textsc{ii}] emission, which shows a weaker enhancement at the center compared to the SE tip.  Peaked central emission in [O~\textsc{i}]$_{63}$ has also been observed by \citet{parkin_2013} in the nucleus of M51, where it was attributed to shocks produced by the Seyfert~2 nucleus.  Cen~A has a strong central active galactic nucleus (AGN); thus, it is possible we see the same type of behaviour in the center as in M51.  This is further supported by the fact that the [O~\textsc{i}]$_{145}$ flux is in good agreement with \citet{2000A&A...355..885U}, who found a total flux of approximately $1.1 \times 10^{-15}$~W~m$^{-2}$ in their center pointing with ISO; however, they find that the [O~\textsc{i}]$_{63}$ flux is $1.96 \times 10^{-14}$~W~m$^{-2}$ in the center alone, with another $5.1 \times 10^{-15}$~W~m$^{-2}$ of flux measured in their south-eastern pointing.  Thus, a large fraction of the total [O~\textsc{i}]$_{63}$ flux likely originates in the nucleus, outside the range of our observations.

\begin{deluxetable}{lcc}
\tabletypesize{\small}
\tablecolumns{3}
\tablecaption{Total integrated flux for cooling lines and infrared continuum in the eastern disk of Cen~A.\label{tbl:int_flux}}
\tablewidth{0pt}
\tablehead{
\colhead{Line} & \colhead{$F$ ($10^{-14}$~W~m$^{-2}$)\tablenotemark{a}}
					& \colhead{Area\tablenotemark{b} ($\sq\arcsec$)}}
  \startdata
 $[$C\,\textsc{ii}]                  & $4.27 \pm 1.28$ & 11232 \\
 $[$N\,\textsc{ii}]$_{122}$          & $0.20 \pm 0.06$ & 9792  \\
 $[$O\,\textsc{i}]$_{63}$            & $1.17 \pm 0.35$ & 11080 \\
 $[$O\,\textsc{i}]$_{145}$           & $0.10 \pm 0.03$ & 10512 \\
 $[$O\,\textsc{iii}]                 & $0.55 \pm 0.17$ & 10656 \\
 $[$N\,\textsc{ii}]$_{205}$          & $0.64 \pm 0.05$ & 22752 \\
 $F_{\mathrm{TIR}}$\tablenotemark{c} & $988 \pm 6$ & 9360 \\
 \enddata
 \tablenotetext{a}{Total integrated flux of each atomic fine structure line we observed for Cen~A.}
 \tablenotetext{b}{The area in square arcseconds over which each total is calculated.  The variations reflect the number of good pixels included in the sum.}
 \tablenotetext{c}{The total integrated flux over the area shown in Figure~\ref{fig:F_tir}, 107568$\sq\arcsec$, is $(3.45 \pm 0.01) \times 10^{-11}$~W~m$^{-2}$.}
\end{deluxetable}

The [N~\textsc{ii}]$_{122}$ and [O~\textsc{iii}] fine structure lines trace ionized gas.  The morphology of the ionized gas, as shown by the [N~\textsc{ii}]$_{122}$ and [O~\textsc{iii}] line emission, is similar to the morphology of the neutral gas and dust emission in the inner half of the map, with sparse detections at greater than a 3$\sigma$ level further out.  While there is an enhancement near the center of the galaxy, the peak of the ionized gas as traced by the [N~\textsc{ii}]$_{122}$ emission is at the SE tip, and is a factor of 2 times greater than the rest of the inner region, and an order of magnitude larger than the outer parts of the map.  In contrast, the peak emission in [O~\textsc{iii}] is coincident with the peak of the [O~\textsc{i}]$_{63}$ and [O~\textsc{i}]$_{145}$.  There are also a few other peaks of emission, one at the SE tip and a little pocket just to the southeast of the SE tip.  The total flux of [O~\textsc{iii}] in our observations is $(5.5 \pm 1.7) \times 10^{-15}$~W~m$^{-2}$, while the total flux in the [N~\textsc{ii}]$_{122}$ line is $(2.0 \pm 0.6) \times 10^{-15}$~W~m$^{-2}$.  \citet{2000A&A...355..885U} find a flux of $7.2 \times 10^{-15}$~W~m$^{-2}$ in [O~\textsc{iii}] for the center pointing, and $1.5 \times 10^{-15}$~W~m$^{-2}$ in [N~\textsc{ii}]$_{122}$ for their center pointing, and an upper limit in their south-east pointing of the same flux.

The area covered by our [N~\textsc{ii}]$_{205}$ map is different than the other five lines we present here, as the observations are centered on the nucleus of the galaxy and do not extend as far east as the PACS maps.  We see that there is a strong detection across the disk, with an emission peak that is roughly a factor of 40 larger than emission detected above and below the plane.  A comparison between the 70~$\mu$m contours and the [N~\textsc{ii}]$_{205}$ line shows that the northwest extra-nuclear peak is also detected in the ionized gas.  The total flux in this map is $(6.4 \pm 0.5) \times 10^{-15}$~W~m$^{-2}$.

\subsection{Line Ratio Diagnostics}\label{subsec:ratio}
\subsubsection{[C~\textsc{ii}], [O~\textsc{i}]$_{63}$, and $F_{\mathrm{TIR}}$ Emission}
The line ratio of ([C~\textsc{ii}]+[O~\textsc{i}]$_{63}$)/$F_{\mathrm{TIR}}$ gives us an indication of the heating efficiency in Cen~A.  The [C~\textsc{ii}] and [O~\textsc{i}]$_{63}$ lines are the dominant coolants in the neutral gas of PDRs.  Thus, their strength tells us how many FUV photons contribute to gas heating, assuming every free electron produced via the photoelectric effect that goes into gas heating eventually results in the emission of one or more [C~\textsc{ii}] or [O~\textsc{i}]$_{63}$ photons.  This value is then divided by the total infrared flux, which indicates how many FUV photons result in dust heating if we assume all dust grains irradiated by FUV flux eventually re-emit infrared continuum emission.

We calculated the total infrared intensity of Cen~A using \emph{Spitzer} MIPS 24~$\mu$m photometry \citep{2012MNRAS.423..197B}, PACS 70 and 160~$\mu$m photometry \citep{2012MNRAS.422.2291P}, and the empirically determined equation for the total infrared intensity (or luminosity) from \citet{2013MNRAS.431.1956G},
\begin{eqnarray}\label{eqn:Ftir}
I_{\mathrm{TIR}} = (2.133 \pm 0.095)\nu_{24} I_{24} + (0.681 \pm 0.028)\nu_{70} I_{70}
	\nonumber \\
	+ (1.125 \pm 0.010) \nu_{160} I_{160}.
\end{eqnarray}
We show a map of the total infrared intensity in Figure~\ref{fig:F_tir}, which has a resolution of 14$\arcsec$.  This map covers the entire disk of Cen~A; however, we only use the region overlapping with our spectroscopic maps for our analysis.  Furthermore, we converted the intensity to flux for comparison to the fine structure line observations.  We achieved this by multiplying the intensity map (in units of W~m$^{-2}$~sr$^{-1}$) by the number of steradians per pixel, to obtain a flux map in units of W~m$^{-2}$~pixel$^{-1}$.  We note that in some studies the far-infrared flux, $F_{\mathrm{FIR}}$, which spans 42~$\mu$m--122~$\mu$m \citep[e.g][]{2008A&A...479..703G}, is used in lieu of the total infrared flux, $F_{\mathrm{TIR}}$ \citep[3~$\mu$m--1100~$\mu$m; ][]{2013MNRAS.431.1956G}.  The two quantities are related via $F_{\mathrm{FIR}} \sim F_{\mathrm{TIR}}/2$ \citep{2001ApJ...549..215D}.

\begin{figure}
\includegraphics[width=\columnwidth]{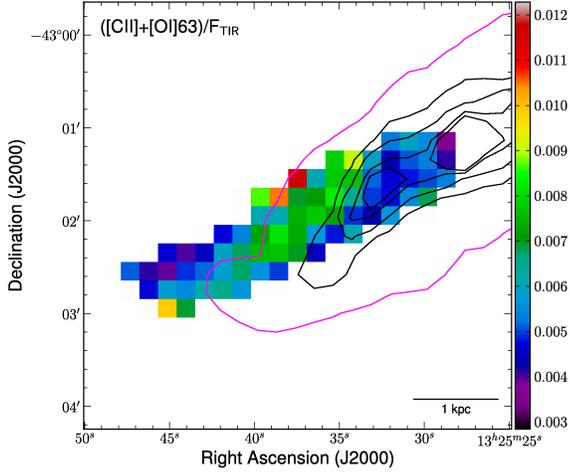}
\caption{Map of the ([C~\textsc{ii}]+[O~\textsc{i}]$_{63}$)/$F_{\mathrm{TIR}}$ line ratio for our observed region of Cen~A.  Contours of the \emph{Herschel} PACS 70~$\mu$m emission are overlaid on top, with the same contour levels as in Figure~\ref{fig:pacs_spec_maps}.}
\label{fig:cena_ratios}
\end{figure}

The heating efficiency map is displayed in Figure~\ref{fig:cena_ratios} and does not show a significant change with increasing radius, with values ranging from $4 \times 10^{-3}$ to $8 \times 10^{-3}$ with an average of $(6 \pm 2) \times 10^{-3}$.  Our value for this ratio is consistent with previous measurements in Cen~A, as \citet{2000A&A...355..885U} find a value of $6 \times 10^{-3}$ in the center and south-east regions.  In other galaxies this ratio typically varies between $10^{-3}$ and 10$^{-2}$ on global scales, as found by \citet{2001ApJ...561..766M}, who studied 60 normal, star-forming galaxies, and \citet{2008ApJS..178..280B}, who conducted an analysis of 227 AGN, starbursts, and normal star-forming galaxies observed with ISO.

\begin{figure}
\includegraphics[width=\columnwidth]{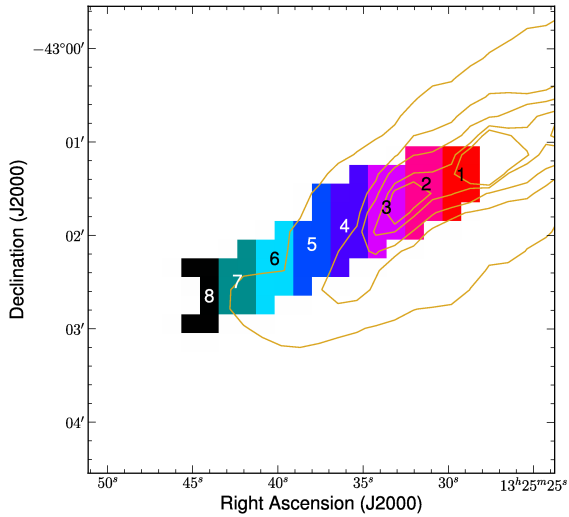}
\caption{A schematic of the radial bins into which we divide our observed line ratio maps.  The colors in this image correspond to the data points in subsequent figures where we use our binned data.  The 70~$\mu$m contours are again overlaid to emphasize the location of this region in the disk.}
\label{fig:schematic}
\end{figure}

We also look at the heating efficiency as a function of infrared color, 70$\mu$m/160$\mu$m (which is often used as a proxy for dust temperature \citep[e.g.][]{2012ApJ...747...81C}).  For this analysis, we have divided the strip of observations used for our analyses here, and in Section~\ref{pdrs}, into eight radial bins as shown by the color-coded schematic in Figure~\ref{fig:schematic}.  Looking at the data in this manner allows us to search for radial variations within these line ratios.  

\begin{figure}
\includegraphics[width=\columnwidth]{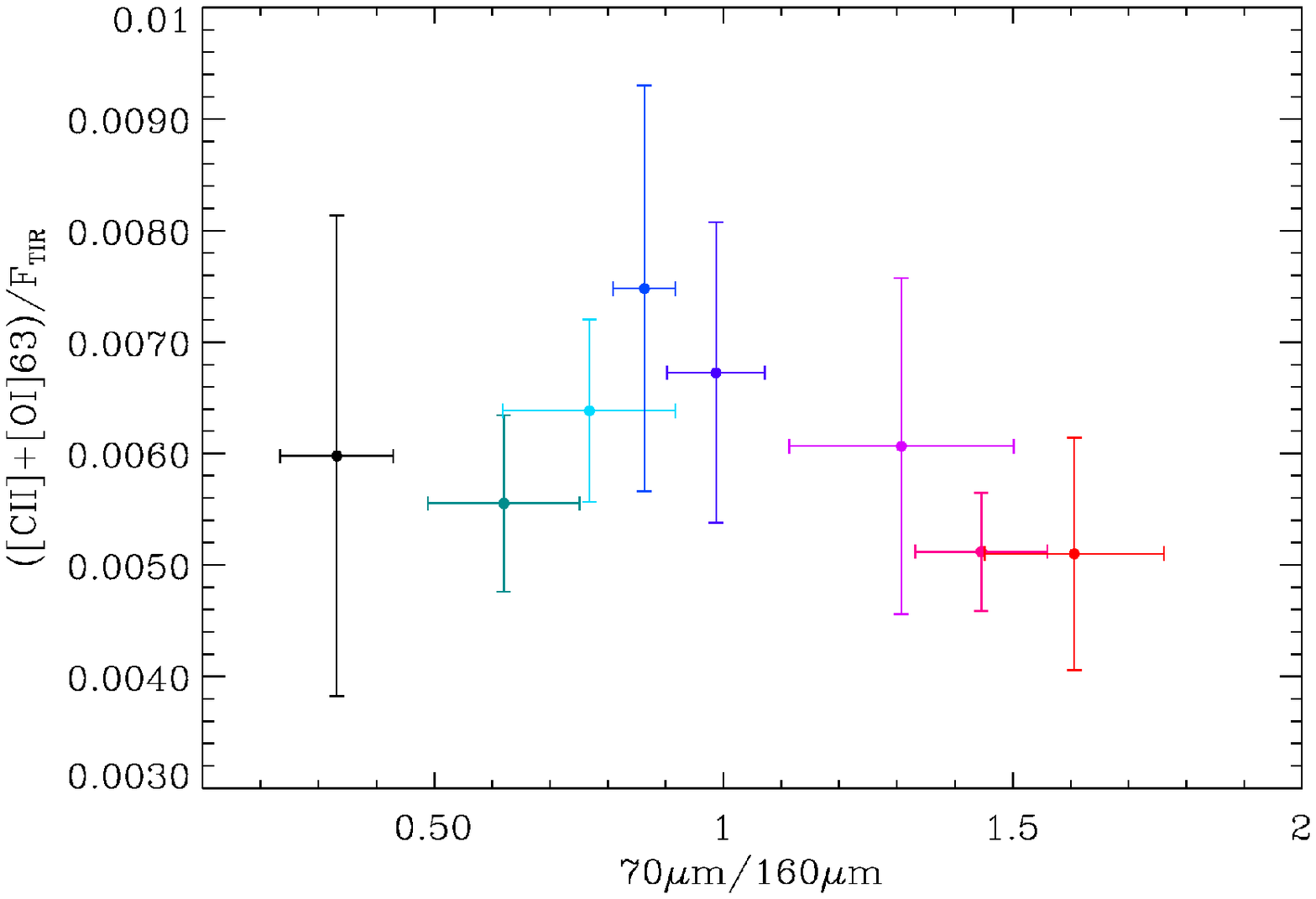}
\includegraphics[width=\columnwidth]{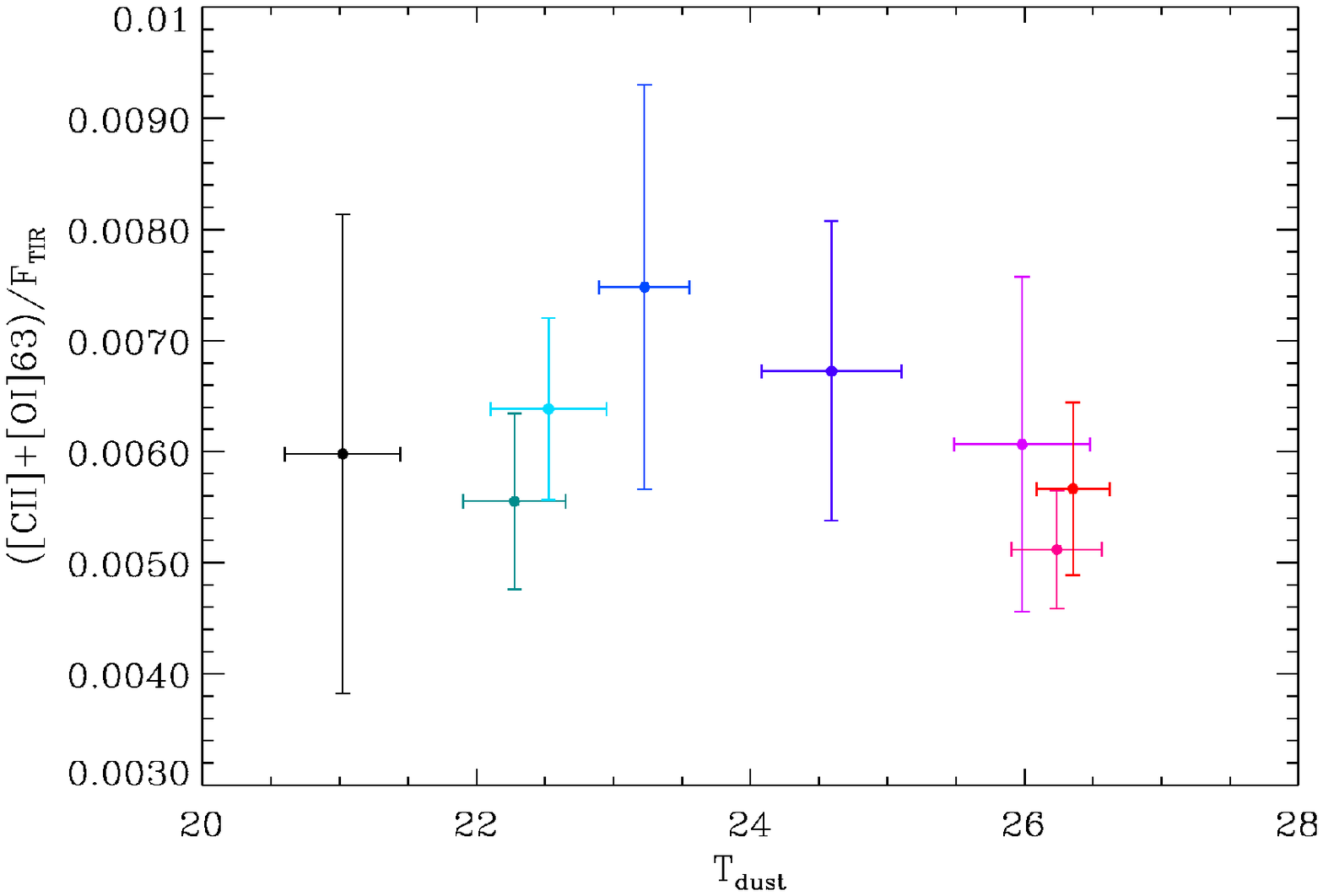}
\caption{The sum of the [C~\textsc{ii}] and [O~\textsc{i}]$_{63}$ cooling lines divided by the total infrared flux, $F_{\mathrm{TIR}}$ for Cen~A.  Each data point represents the average value within each bin displayed in Figure~\ref{fig:schematic}.  Systematic uncertainties due to calibration are not shown. \emph{top}: The ([C~\textsc{ii}]+[O~\textsc{i}]$_{63}$)/$F_{\mathrm{TIR}}$ line ratio plotted as a function of the 70$\mu$m/160$\mu$m color.  \emph{bottom}: The ([C~\textsc{ii}]+[O~\textsc{i}]$_{63}$)/$F_{\mathrm{TIR}}$ line ratio plotted as a function of dust temperature. No significant radial trend in the ratio is seen.}
\label{fig:heat_eff}
\end{figure}

A plot of the heating efficiency as a function of the 70$\mu$m/160$\mu$m color is shown in the top panel of Figure~\ref{fig:heat_eff}.  Each point represents the average value in each bin, while uncertainties are estimated from the standard deviations of the quantities in each bin.  The innermost bin (shown in red) has a value of $\sim 5 \times 10^{-3}$, we see an increase in the middle bins up to almost $8 \times 10^{-3}$ (shown in blue), then a decrease in the outermost bins.  In the bottom panel of Figure~\ref{fig:heat_eff}, we show a plot of the heating efficiency as a function of dust temperature, determined by \citet{2012MNRAS.422.2291P}.  The innermost bins show the warmest dust.  In addition, we do not see a significant trend with increasing infrared color within uncertainties for either parameter space.  This is an interesting result because on global scales, \citet{2001ApJ...561..766M} observe a reduced heating efficiency in galaxies with the warmest 60~$\mu$m/100~$\mu$m colors, and \citet{2008ApJS..178..280B} found no overall decreasing trend in heating efficiency with increasing dust temperatures.  However, a decrease in heating efficiency with increasing infrared color (and thus dust temperature) has been previously observed within individual galaxies by \citet{2012A&A...548A..91L} in an H~\textsc{ii} region within the Large Magellanic Cloud, by \citet{2012ApJ...747...81C} in NGC~1097 and NGC~4559, and by \citet{parkin_2013} in M51.  A suppression in heating efficency has been attributed to dust grains becoming too positively charged for the photoelectric effect to efficiently free electrons \citep[e.g.][]{2001ApJ...561..766M}.  Thus, for the observed region of Cen~A's disk, the dust grains are largely unaffected by the impinging radiation field.

\subsubsection{Molecular Gas Cooling}
\begin{figure}
\includegraphics[width=\columnwidth]{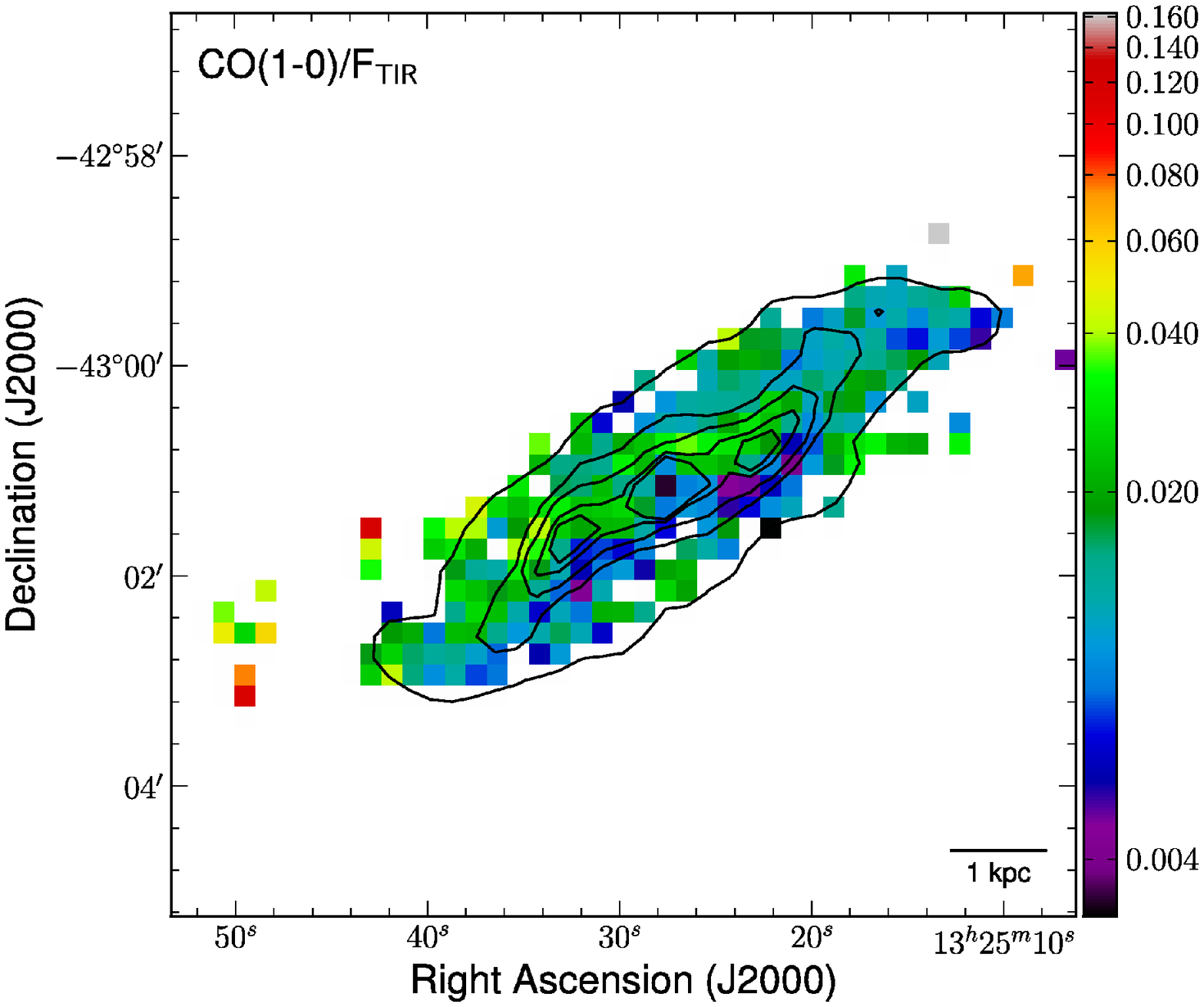}
\includegraphics[width=\columnwidth]{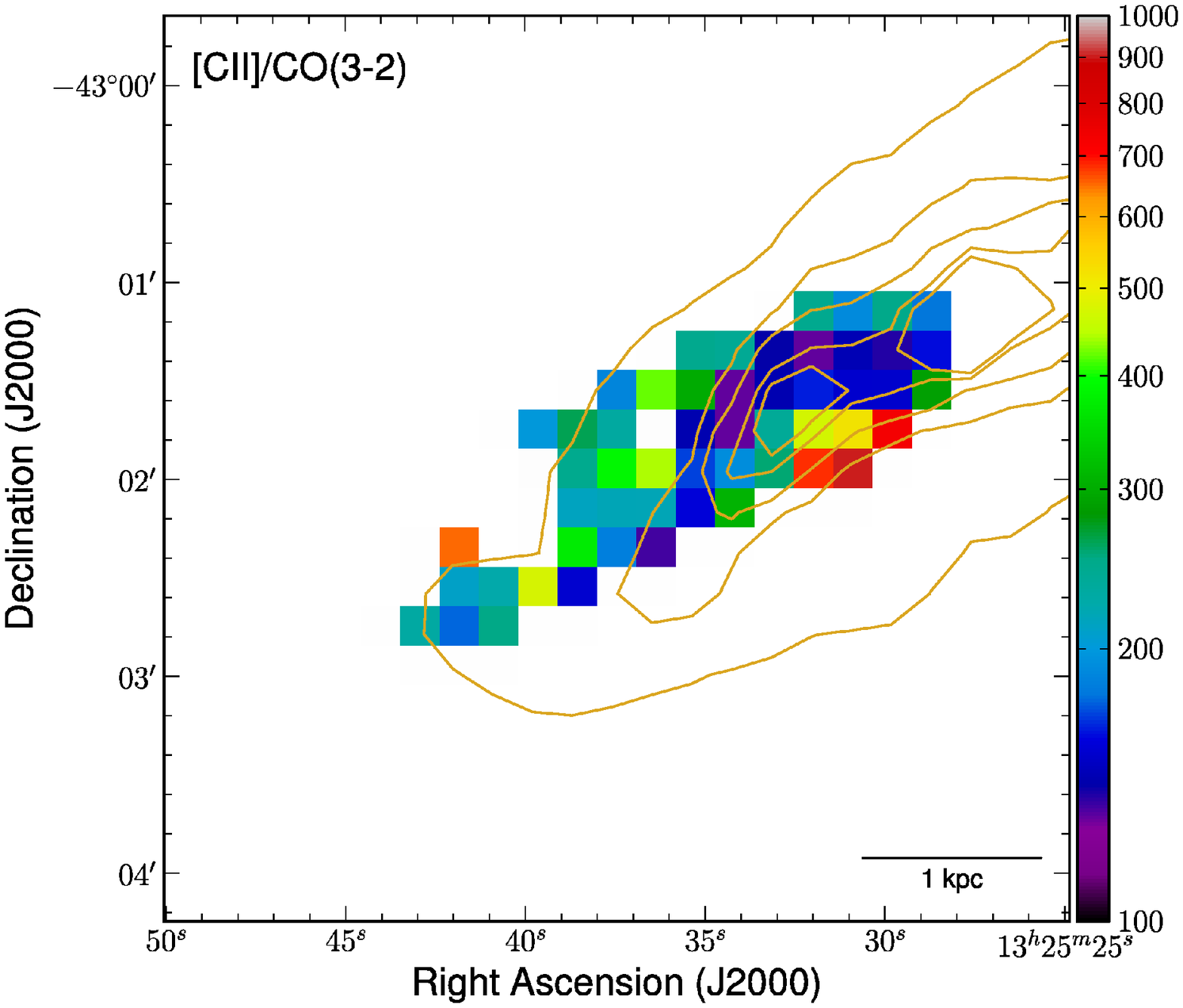}
\caption{Maps of the CO($J=1-0$)/$F_{\mathrm{TIR}}$ (\emph{top}) and [C~\textsc{ii}]/CO($J=3-2$) (\emph{bottom}) line ratios for our observed region of Cen~A.  Contours of the \emph{Herschel} PACS 70~$\mu$m emission showing the warm dust distribution are overlaid on top, with the same contour levels as in Figure~\ref{fig:pacs_spec_maps}.  When reading the colourscale for the CO($J=1-0$)/$F_{\mathrm{TIR}}$ line ratio the numbers should be multiplied by $10^{-4}$.}
\label{fig:cena_co_ratios}
\end{figure}

CO also contributes to the cooling via its rotational lines, although its contribution is small in comparison to the [C\,\textsc{ii}] and [O\,\textsc{i}]$_{63}$ cooling lines.  Utilizing the CO($J=1-0$) integrated intensities reported at three positions in the disk of Cen~A from \citet{1990ApJ...363..451E}, and measuring the corresponding CO($J=3-2$) integrated intensities in our map, we find an average CO($J=3-2$)/CO($J=1-0$) ratio of $0.42 \pm 0.04$ ($11 \pm 1$) when the CO integrated intensities are in units of K~km~s$^{-1}$ (W~m$^{-2}$).  Dividing the CO($J=3-2$) map by the average CO($J=3-2$)/CO($J=1-0$) ratio we obtain an estimate of the CO($J=1-0$) distribution.  In the top panel of Figure~\ref{fig:cena_co_ratios} we show the estimated line ratio CO($J=1-0$)/$F_{\mathrm{TIR}}$ for Cen~A.  Similar to the ([C~\textsc{ii}]+[O~\textsc{i}]$_{63}$)/$F_{\mathrm{TIR}}$ line ratio, we do not detect a significant trend with increasing radius in this line ratio.  Over the area outlined in Figure~\ref{fig:schematic} (for comparison with other line ratios) we find an average CO($J=1-0$)/$F_{\mathrm{TIR}}$ ratio of $(1.9 \pm 0.2) \times 10^{-6}$.

The average value of the [C~\textsc{ii}]/CO($J=3-2$) line ratio (shown in the bottom panel of Figure~\ref{fig:cena_co_ratios}) is $(2.7 \pm 1.7) \times 10^{2}$, while the corresponding average value of the [C~\textsc{ii}]/CO($J=1-0$) line ratio is $(3 \pm 2) \times 10^{3}$ across the strip.  This second value agrees within uncertainties with the average found for a sample of starburst galaxies and Galactic star forming regions including Cen~A, which is 4200 (after dividing their CO integrated intensities by the main beam efficiency) \citep{1991ApJ...373..423S}.  They find [C~\textsc{ii}]/CO($J=1-0$) to be roughly 4070 for Cen~A in particular.

Converting the CO flux to a molecular hydrogen mass we can compare the line/$F_{\mathrm{FIR}}$ ratios with recent results from \citet{2011ApJ...728L...7G}.  They investigated the parameter space of line/$F_{\mathrm{FIR}}$ vs. $L_{\mathrm{FIR}}$/$M_{\mathrm{H}_{2}}$ for a subset of the SHINING sample of galaxies.  The ratio $L_{\mathrm{FIR}}$/$M_{\mathrm{H}_{2}}$ represents the number of stars formed per unit mass of molecular gas per unit of time.  We convert our CO($J=3-2$) integrated intensity to an H$_{2}$ mass assuming an X$_{\mathrm{CO}}$ factor of $(2 \pm 1) \times 10^{20}$~cm$^{-2}$~(K~km~s$^{-1}$)$^{-1}$, typical for the Milky Way \citep{1988A&A...207....1S}, and the CO($J=3-2$)/CO($J=1-0$) ratio of $0.42 \pm 0.04$, calculated as described above.  We do not have global measurements for the various fine structure lines, thus we opt to instead measure the average $L_{\mathrm{FIR}}$/$M_{\mathrm{H}_{2}}$ ratio over the area shown in Figure~\ref{fig:schematic}.  We list the average $L_{\mathrm{FIR}}$/$M_{\mathrm{H}_{2}}$ ratio along with the average line/$F_{\mathrm{FIR}}$ ratios for Cen~A in Table~\ref{tbl:line_LonMH2}.  These line/$F_{\mathrm{FIR}}$ ratios are consistent with the range of values for the sample of various galaxy types from \citet{2011ApJ...728L...7G}.  The average value of $L_{\mathrm{FIR}}$/$M_{\mathrm{H}_{2}}$, $(5.4 \pm 0.2)$~L$_{\odot}$~M$_{\odot}^{-1}$, is consistent with galaxies with the lowest values of $L_{\mathrm{FIR}}$/$M_{\mathrm{H}_{2}}$.  A search for radial trends by plotting the line/$F_{\mathrm{FIR}}$ ratios as a function of $L_{\mathrm{FIR}}$/$M_{\mathrm{H}_{2}}$ for each bin shows no significant radial trends.  It is possible we do not see a trend because we do not probe to high enough scales to see the deficit take effect at $L_{\mathrm{FIR}}$/$M_{\mathrm{H}_{2}} \gtrsim 80$~L$_{\odot}$~M$_{\odot}^{-1}$ as shown by \citet{2011ApJ...728L...7G}.

\begin{deluxetable}{lcc}
\tabletypesize{\small}
\tablecolumns{2}
\tablecaption{$L_{\mathrm{FIR}}$/$M_{\mathrm{H}2}$ and line/$F_{\mathrm{FIR}}$ for the Eastern Disk of Cen~A.\label{tbl:line_LonMH2}}
\tablewidth{0pt}
\tablehead{
\colhead{Line} & \colhead{Average}}
  \startdata
  \multicolumn{2}{c}{line/$F_{\mathrm{FIR}}$ ($10^{-4}$)} \\
  \hline
  $[$C\,\textsc{ii}]         & $84 \pm 30$   \\
  $[$N\,\textsc{ii}]$_{122}$ & $4 \pm 1$     \\
  $[$O\,\textsc{i}]$_{63}$   & $22 \pm 7$    \\
  $[$O\,\textsc{i}]$_{145}$  & $2.0 \pm 0.6$ \\
  $[$O\,\textsc{iii}]        & $10 \pm 3$    \\
  \hline
  \multicolumn{2}{c}{L$_{\odot}$~M$_{\odot}^{-1}$} \\
  \hline
  $L_{\mathrm{FIR}}$/$M_{\mathrm{H}2}$ & $5.4 \pm 0.2$  \\
  \enddata
\end{deluxetable}

\subsubsection{Ionized Gas}\label{oiiionnii}
\begin{figure}
\includegraphics[width=\columnwidth]{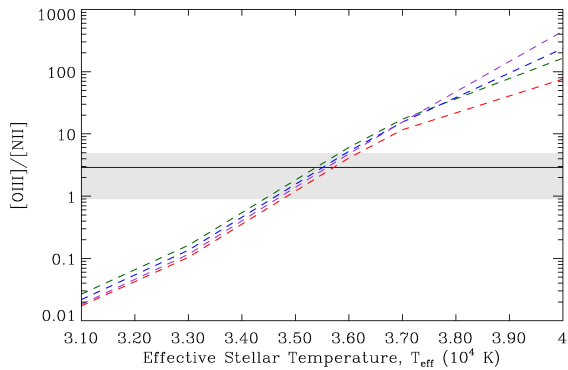}
\caption{A comparison of the average observed [O~\textsc{iii}]/[N~\textsc{ii}]$_{122}$ line ratio for CenA to predicted line ratios as a function of effective stellar temperature for models of an H~\textsc{ii} region. The black solid line represents the global average over our observed line ratio, while the shaded region encompasses the range of values within uncertainty.  The dashed lines show the predicted line ratio for gas densities of $10^{2}$ (red), $10^{3}$ (green), $10^{4}$ (blue), and $10^{5}$~cm$^{-3}$ (purple) using the H~\textsc{ii} region model of \citet{1985ApJS...57..349R}.}
\label{fig:OIIIonNII}
\end{figure}

The observed [O~\textsc{iii}]/[N~\textsc{ii}]$_{122}$ line ratio has been used in high redshift sources to interpret either the strength of the ionization parameter, $U$, which is the number of ionizing photons divided by the gas density within the narrow line region of an AGN \citep{2009ApJ...701.1147A}, or the stellar classification of the youngest stars in an H~\textsc{ii} region, depending on the type of region one is investigating \citep{2011ApJ...740L..29F}.  We have taken the average observed [O~\textsc{iii}]/[N~\textsc{ii}]$_{122}$ line ratio for the disk of Cen~A and plotted it in Figure~\ref{fig:OIIIonNII} (black solid line), with the shaded region outlining the range of values within calibration uncertainties.  We opt to plot the average rather than the binned values because we do not observe significant variations in the [O~\textsc{iii}]/[N~\textsc{ii}]$_{122}$ line ratio map.  Overlaid on the observed ratio are predicted line ratios as a function of stellar temperature from the H~\textsc{ii} region models of \citet{1985ApJS...57..349R}, where the red, green, blue and purple dashed lines represent gas densities of $10^{2}$, $10^{3}$, $10^{4}$ and $10^{5}$~cm$^{-3}$, respectively.  The [O~\textsc{iii}]/[N~\textsc{ii}]$_{122}$ line ratio we derive falls within a range of stellar effective temperature of approximately $3.45 \times 10^{4}$ and $3.62 \times 10^{4}$~K, which corresponds to stellar classifications of O9.5 or O9 \citep{1996ApJ...460..914V}.  However, we note that if the AGN contributes in part to the observed emission, the stellar classifications will shift to later types.

We have chosen not to probe how [O~\textsc{iii}]/[N~\textsc{ii}]$_{122}$ varies as a function of $U$, because we are investigating the disk of Cen~A rather than the nucleus.  However, if the AGN were to contribute partially to the observed emission in the center, it might explain why our observed [O~\textsc{iii}]/[N~\textsc{ii}]$_{122}$ line ratio is higher (and thus the stellar classification is earlier) than that observed in M51 \citep{parkin_2013}.

\section{PDR modelling}\label{pdrs}

\subsection{Correcting the [C~{\footnotesize II}] emission for the Ionized Gas Contribution}\label{ion_gas}
The emission in the [C~\textsc{ii}] line comes from three sources: dense neutral gas, ionized gas, and diffuse neutral gas.  For us to properly utilize the photodissociation region model in Section~\ref{pdr_model} to interpret our diagnostic far-infrared spectral lines, we need to isolate the [C~\textsc{ii}] emission associated with the dense, neutral gas found in PDRs.  The ionized gas contribution can be determined by comparing two observed line ratios, namely the [N~\textsc{ii}]$_{122}$/[N~\textsc{ii}]$_{205}$ and [C~\textsc{ii}]/[N~\textsc{ii}]$_{205}$ ratios, to a theoretical prediction for each line as a function of electron density in an H~\textsc{ii} region \citep[e.g.][]{2006ApJ...652L.125O, parkin_2013}.  The low critical density of the [N~\textsc{ii}]$_{205}$ line (approximately 44~cm$^{-3}$ at $T_{e}=8000$~K) implies that emission via this transition can originate in diffuse ionized gas as well as in higher density ionized gas such as is typically seen in H~\textsc{ii} regions.  However, the critical density of the [C~\textsc{ii}] line is almost identical (46~cm$^{-3}$) to that of the [N~\textsc{ii}]$_{205}$ line, thus both lines will probe ionized gas of the same density, which is key for using this method to remove the ionized gas contribution to the observed [C~\textsc{ii}] emission.

To calculate the level populations (and thus the predicted fluxes) for the two [N~\textsc{ii}] transitions we employ the Einstein coefficients from \citet{1997A&AS..123..159G} and collision strengths for collisions with electrons from \citet{2004MNRAS.348.1275H}.  For the [C~\textsc{ii}] line level populations we use the Einstein coefficients of \citet{1998A&AS..131..499G} and collision strengths of 
\citet{1992ApJS...80..425B}.  Due to the lack of accurate measurements of the gas phase abundances of C or N, as well as the metallicity in Cen~A, we adopt Solar gas phase abundances and assume no metallicity gradient within the region we are investigating.   The abundances we choose are from \citet{1996ARA&A..34..279S}, and are C/H~$= 1.4 \times 10^{-4}$ and N/H~$= 7.9 \times 10^{-5}$.

Our observed [N~\textsc{ii}]$_{122}$/[N~\textsc{ii}]$_{205}$ line ratio is initially calculated for the small region of overlap between the observations of the two lines.  We convolve the [N~\textsc{ii}]$_{122}$ map to the resolution of the [N~\textsc{ii}]$_{205}$ map ($17\arcsec$) using a Gaussian kernel, then align them to a common pixel grid.  Next, we convert the units of the [N~\textsc{ii}]$_{122}$ map to match those of the [N~\textsc{ii}]$_{205}$ (Jy~Hz~beam$^{-1}$) and then calculate the line ratio in each of the overlapping pixels.

Using a theoretical curve of the [N~\textsc{ii}]$_{122}$/[N~\textsc{ii}]$_{205}$ line ratio as a function of electron density for an H~\textsc{ii} region, we determine the electron density at which our observed [N~\textsc{ii}]$_{122}$/[N~\textsc{ii}]$_{205}$ ratio matches that of the theoretical prediction for each pixel in our line ratio map.  We find a mean electron density of 6.3~cm$^{-3}$ with lower and upper limits of 0.8 and 12.3~cm$^{-3}$.  Given that there is little overlap between our [N~\textsc{ii}]$_{205}$ map and our [N~\textsc{ii}]$_{122}$ map, we choose to take the mean observed ratio and standard deviation as the adopted [N~\textsc{ii}]$_{122}$/[N~\textsc{ii}]$_{205}$ measurement for the full area of our PACS observations.  Thus, we find [N~\textsc{ii}]$_{122}$/[N~\textsc{ii}]$_{205} = 0.8 \pm 0.2$.

To confirm the low gas density that we find using the [C~\textsc{ii}], [N~\textsc{ii}]$_{122}$, and [N~\textsc{ii}]$_{205}$ lines, we have also looked at the [S~\textsc{iii}](18.71~$\mu$m)/[S~\textsc{iii}](33.48~$\mu$m) line ratio, which is sensitive primarily for gas densities between 10$^{2}$~cm$^{-3}$ and 10$^{4}$~cm$^{-3}$ \citep{2007ApJ...669..269S}.  Through the \emph{Spitzer} Heritage Archive we obtained observations of Cen~A taken with the \emph{Spitzer's} Infrared Spectrograph (IRS) from three different AORs (4939776, 4939264, and 8767488), two of which are centered on the nucleus and one of which looks at a small region in the disk.  The data were processed with the SMART package \citep{2010PASP..122..231L}, then projected using the CUBISM package \citep{2007PASP..119.1133S}.  In all three cases we find that the [S~\textsc{iii}](18.71~$\mu$m)/[S~\textsc{iii}](33.48~$\mu$m) is less than roughly 0.5, which implies an ionized gas density of less than 10$^{2}$~cm$^{-3}$ \citep{2007ApJ...669..269S, 2009ApJ...693..713S}.  However, this line ratio becomes insensitive to densities below approximately 10$^{2}$, so we cannot be more specific about the ionized gas density using just the [S~\textsc{iii}] lines.  Nonetheless, these results are consistent with the densities derived above using the [C~\textsc{ii}], [N~\textsc{ii}]$_{122}$, and [N~\textsc{ii}]$_{205}$ lines, thus providing confidence in the ionized gas density we find.

Based on our estimate of the electron density, we then determine the theoretical prediction for the [C~\textsc{ii}]/[N~\textsc{ii}]$_{205}$ ratio in ionized gas.  A map of [C~\textsc{ii}] emission originating in ionized gas is then generated from the predicted [C~\textsc{ii}]/[N~\textsc{ii}]$_{205}$ ratio, our [N~\textsc{ii}]$_{122}$ map, and our assumed constant [N~\textsc{ii}]$_{122}$/[N~\textsc{ii}]$_{205}$ line ratio.  For comparison with the PDR model in Section~\ref{pdr_model}, we remove the fraction of [C~\textsc{ii}] emission originating in ionized gas from the total observed emission, which in general is quite low.  The majority of our map demonstrates a contribution of roughly 10 to 20\%, with the pixels showing the highest ionized gas contribution falling at the edge of the map farthest from the center of the galaxy.

\subsection{Potential Non-PDR Contributions to the Observed [C~II] and $F_{\mathrm{TIR}}$ Emission}
We note that there is a possibility that, of the [C~\textsc{ii}] emission stemming from neutral gas, some may come from the diffuse ISM rather than from PDRs.  However, \citet{2000A&A...355..885U} looked into this possibility and concluded that less than 5\% of the total [C~\textsc{ii}] emission originated in non-PDR, diffuse gas within their ISO observations of Cen~A.  Furthermore, we calculate the ratio of H$_{2}$/H~\textsc{i} using the maps from \citet{2012MNRAS.422.2291P} and find an average value of roughly 5 through the area covered by our spectroscopic strips, suggesting that the gas is H$_{2}$ dominated.

In addition, it is possible that not all of the observed $F_{\mathrm{TIR}}$ emission stems from PDRs as well; for example, a fraction could come from H~\textsc{ii} regions.  However, in the context of the PDR model considered here (see below), we find that it is unlikely that a significant fraction of the observed $F_{\mathrm{TIR}}$ emission in Cen~A originates in low-intensity PDRs or diffuse gas.  We note that we do not take these contributions into account in the following analysis.

\subsection{The Model}\label{pdr_model}
A comparison between observed line ratios and a PDR model allows us to diagnose the physical properties of the PDRs from which the fine structure line emission originates.  Here we choose to use the PDR model of \citet{1999ApJ...527..795K,2006ApJ...644..283K}, which has been updated and expanded from the model of \citet{1985ApJ...291..722T}.  This particular model assumes the PDR is a plane-parallel, semi-infinite slab and is only parameterized by two free variables: the hydrogen gas density, $n$, and the strength of the impinging far-ultraviolet (FUV) radiation field normalized to the Habing field \citep[$1.6 \times 10^{-3}$~erg~cm$^{-2}$~s$^{-1}$; ][]{1968BAN....19..421H}, $G_{0}$.  The model simultaneously treats the thermal balance, chemical network and radiative transfer and produces a grid of predicted fine structure line strengths as a function of $n$ and $G_{0}$.  By comparing observed line ratios to the predicted ones we can extract the corresponding $n$ and $G_{0}$.

\begin{figure}
\includegraphics[width=0.85\columnwidth]{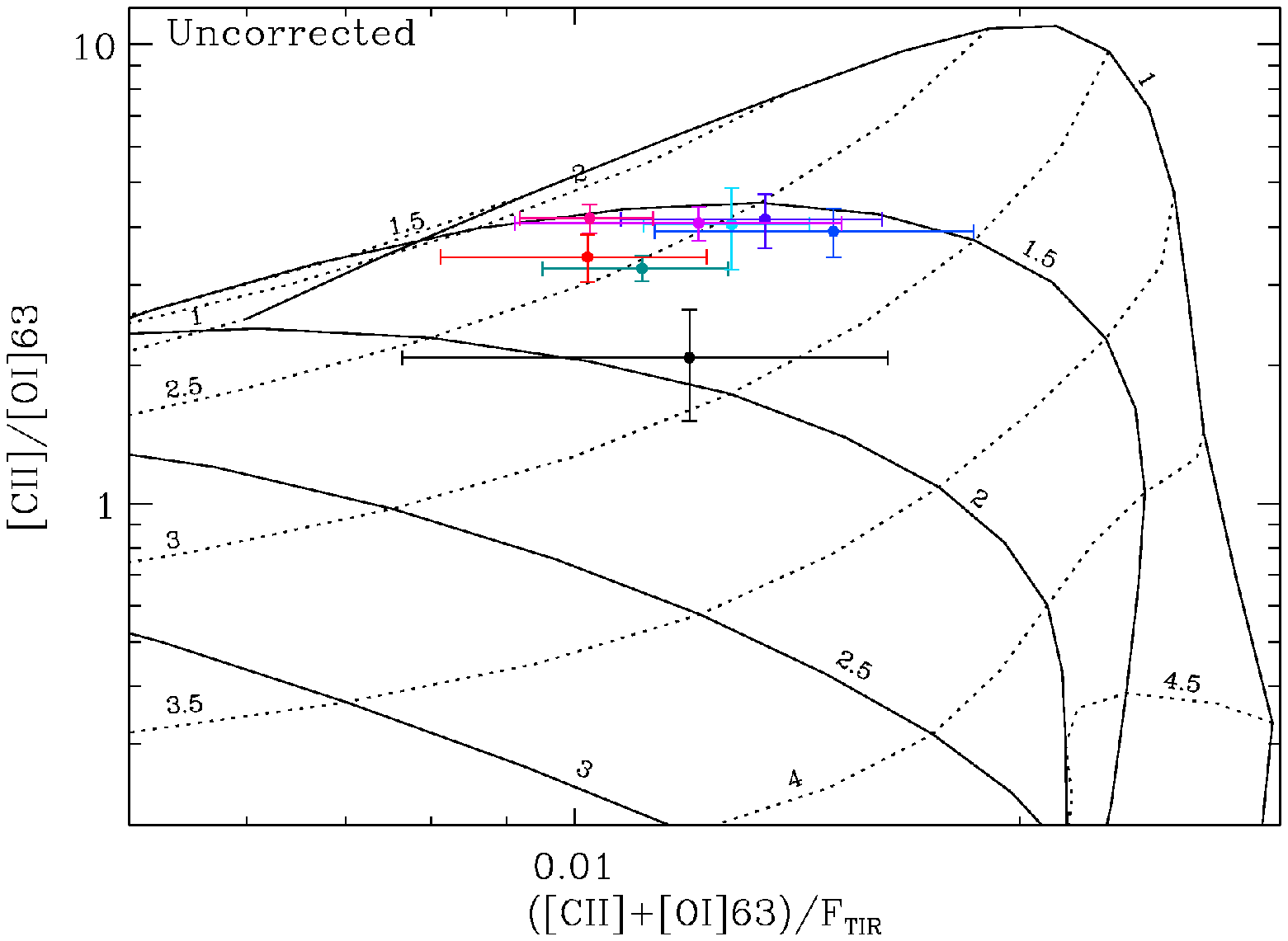}
\includegraphics[width=0.85\columnwidth]{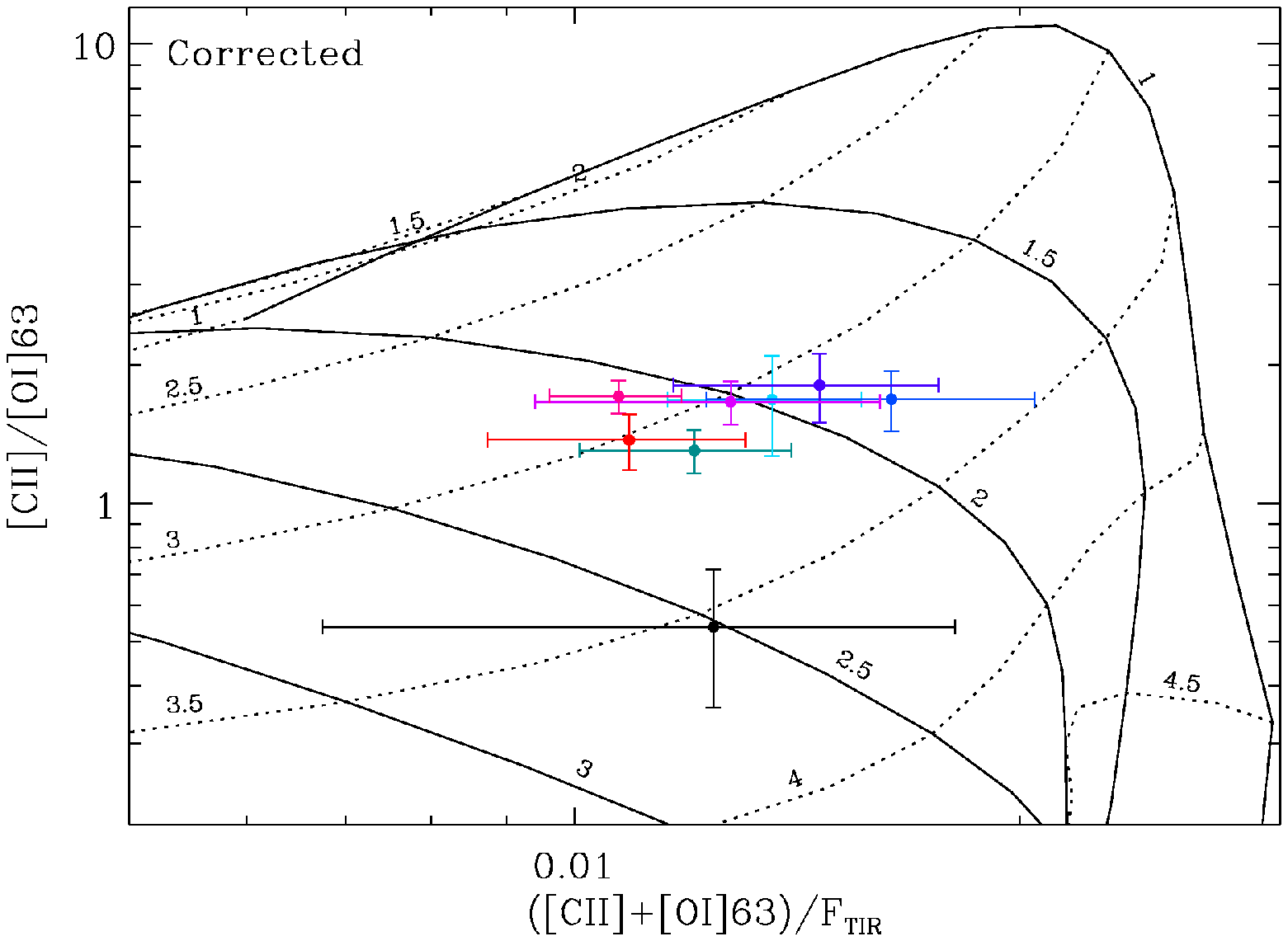}
\includegraphics[width=0.85\columnwidth]{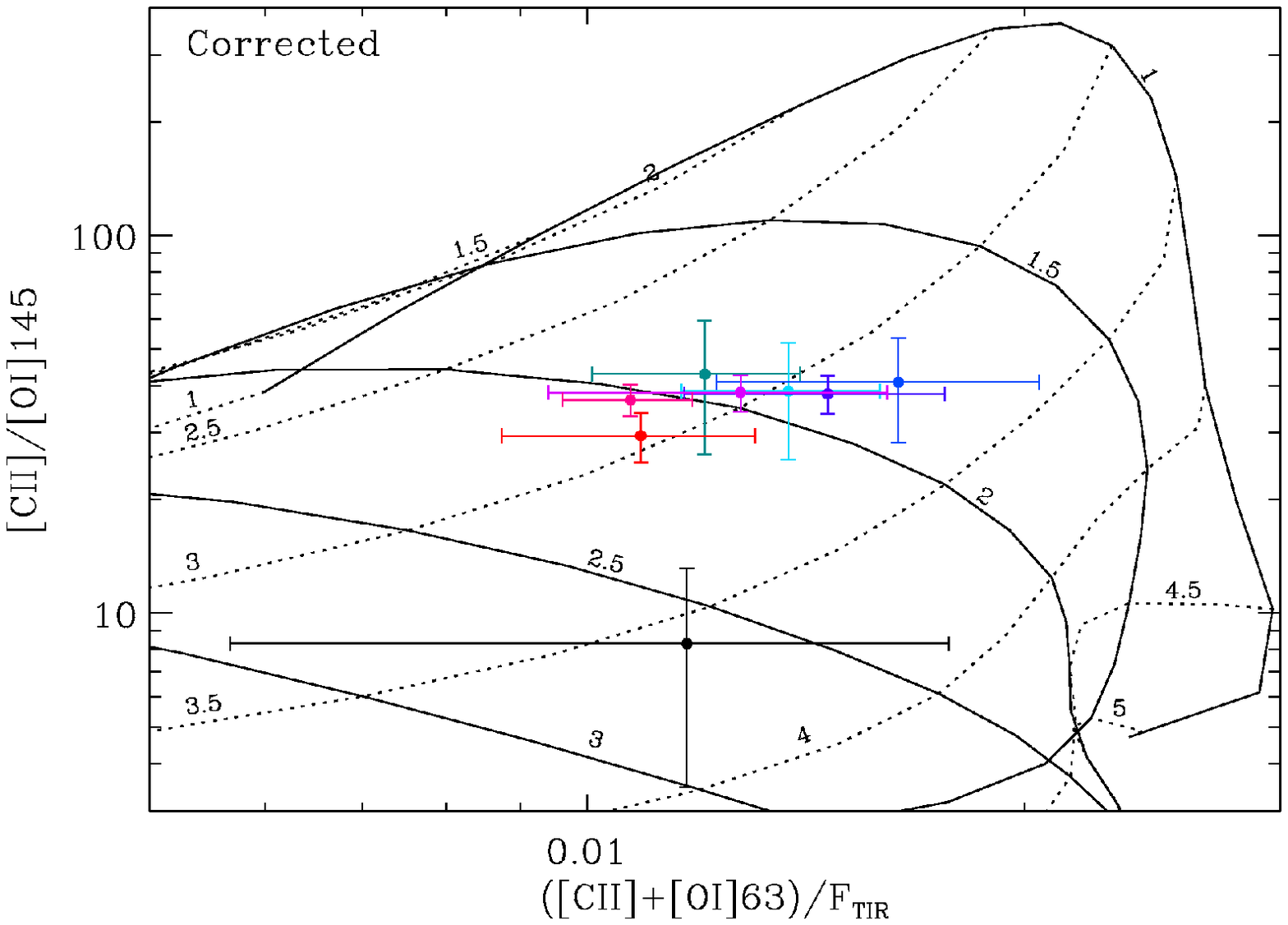}
\caption{\emph{top}: A comparison of our observed (and uncorrected) [C~\textsc{ii}]/[O~\textsc{i}]$_{63}$ and ([C~\textsc{ii}]+[O~\textsc{i}]$_{63}$)/$F_{\mathrm{TIR}}$ line ratios to the PDR model of \citet{1999ApJ...527..795K} for the eastern half of Centaurus~A. The data points represent our observations, with each color corresponding to a radial bin as shown in the top left panel.  The solid lines represent contours of constant log$G_{0}$ while the dotted lines represent curves of constant log($n$/cm$^{-3}$).  \emph{middle}: The same plot as in the top right panel, but here the [C~\textsc{ii}] and [O~\textsc{i}]$_{63}$ have been corrected as described in the text. \emph{bottom}: Our observed [C~\textsc{ii}]/[O~\textsc{i}]$_{145}$ and ([C~\textsc{ii}]+[O~\textsc{i}]$_{63}$)/$F_{\mathrm{TIR}}$ line ratios compared to the PDR model.  The lines and data points are the same as in the top right and bottom left panels.}
\label{fig:pdr_plots}
\end{figure}

For our investigation we choose to utilize the line ratio parameter space of [C~\textsc{ii}]/[O~\textsc{i}]$_{63}$ vs. ([C~\textsc{ii}]+[O~\textsc{i}]$_{63}$)/$F_{\mathrm{TIR}}$.  In order to search for radial variations in the disk of Cen~A we have again divided our observed line ratio maps into eight bins as shown in Figure~\ref{fig:schematic} to measure $n$ and $G_{0}$.    The unweighted average observed values in each bin are overlaid on the PDR model grid in the top panel of Figure~\ref{fig:pdr_plots}, with the error bars incorporating both the measurement uncertainties of the observations as well as the standard deviation of the data in each bin.  Even if the observed total flux emits from both the near and far sides of the cloud(s) because the cloud is optically thin to dust continuum infrared photons, the model assumes it only emits from the side exposed to the source of FUV flux.  Thus, we divide the total infrared flux, $F_{\mathrm{TIR}}$, by a factor of two as recommended by \citet{1999ApJ...527..795K}.   The resulting values of $n$, $G_{0}$ and the temperature at the surface of the PDR, $T$, are presented in Table~\ref{table:cena_pdr_model_results} under the ``Uncorrected" heading.

The PDR model assumes the [C~\textsc{ii}] emission originates only in neutral gas, but as described in Section~\ref{ion_gas}, [C~\textsc{ii}] emission can be produced in both neutral and ionized gas.  Thus, to properly compare our observations to the model we need to remove the 10-20\% contribution from the ionized gas.  We also need to make a correction to the [O~\textsc{i}]$_{63}$ observations that stems from geometrical effects of many PDRs in a given observation for extragalactic sources.  We see PDRs at all orientations with respect to our line of sight, but when a spectral line is optically thick and the cloud is lit from behind, we will not observe emission from that line, as is the case for the [O~\textsc{i}]$_{63}$ line. \citet{1999ApJ...527..795K} state that as a result of the optically thick line and various PDR orientations, we only observe about half of the total [O~\textsc{i}]$_{63}$ emission produced, while the remaining half radiates away from the line of sight.  Following the recommendation of \citet{1999ApJ...527..795K}, we increase our observed [O~\textsc{i}]$_{63}$ flux by a factor of 2, as we have previously done with M51 \citep{parkin_2013}.  We caution the reader that there is some uncertainty in this correction factor that should be kept in mind when interpreting the following results.  We show the fully corrected line ratios compared to the PDR model in the middle panel of Figure~\ref{fig:pdr_plots} and tabulate the results in Table~\ref{table:cena_pdr_model_results}.  We see that with these changes the data points shift down and slightly to the right, corresponding to increases in both $G_{0}$ and $n$.

Our assumption that the [O~\textsc{i}]$_{63}$ line is optically thick while the [O~\textsc{i}]$_{145}$ and [C~\textsc{ii}] lines are optically thin can be supported by comparing the observed [O~\textsc{i}]$_{63}$/[O~\textsc{i}]$_{145}$ line ratio to theoretical curves representing the line ratio as a function of temperature and density.  In Cen~A we determine an average [O~\textsc{i}]$_{63}$/[O~\textsc{i}]$_{145}$ line ratio of $11.6 \pm 3.5$.  Comparing this value to the theoretical curves of \citet{2006A&A...446..561L} we find that for $100 \lesssim T \lesssim 2000$~K the observed line ratio is higher than would be the case if both [O~\textsc{i}] lines are optically thick, and lower than predicted if both [O~\textsc{i}] lines are optically thin and $n \lesssim 10^{3}$~cm$^{-2}$.  Therefore, it is likely that the [O~\textsc{i}]$_{63}$ line is optically thick while the [O~\textsc{i}]$_{145}$ is optically thin.

As a consistency check, in Figure~\ref{fig:pdr_plots} (bottom panel) we also show a comparison of our observations with the PDR model parameter space [C~\textsc{ii}]/[O~\textsc{i}]$_{145}$ vs. ([C~\textsc{ii}]+[O~\textsc{i}]$_{63}$)/$F_{\mathrm{TIR}}$, which has the advantage of the optically thin [O~\textsc{i}]$_{145}$ line.  We find that this parameter space gives consistent solutions for $n$ and $G_{0}$ as those derived from the plot in the bottom left panel of Figure~\ref{fig:pdr_plots}.  This further suggests to us that our assumption that the [O~\textsc{i}]$_{63}$ line is optically thick is valid.

\begin{deluxetable}{lcccccc}
\tabletypesize{\small}
\tablecolumns{7}
\tablecaption{Properties of the gas derived from the PDR model\label{table:cena_pdr_model_results}}
\tablewidth{0pt}
\tablehead{
\colhead{Case}
 & \colhead{Bin} & Angular Distance\tablenotemark{a} ($'$) & Area ($\sq\arcsec$) & \colhead{log($n$/cm$^{-3}$)} & \colhead{log$G_{0}$}
	& \colhead{$T$ (K)}}
 \startdata
 uncorrected\tablenotemark{b}
 & 1 & 0.33 & 1008 & (2.25--2.5)$^{+0.5}_{-0.75}$ & (1.5--1.75)$^{+0.5}_{-0.25}$
 	& (120--170)$^{+655}_{-40}$ \\
 & 2 & 0.75 & 1296 & (2.0--2.25)$^{+0.5}_{-0.5}$  & (1.5--1.75)$^{+0.25}_{-0.5}$
 	& (140--210)$^{+410}_{-70}$ \\
 & 3 & 1.2 & 1296 & (2.25--2.5)$^{+0.5}_{-0.75}$ & (1.5--1.75)$^{+0.25}_{-0.25}$
 	& (120--170)$^{+450}_{-40}$ \\
 & 4 & 1.67 & 1440 & (2.5--2.75)$^{+0.5}_{-1.0}$  & (1.5--1.75)$^{+0.25}_{-0.25}$
 	& (110--150)$^{+470}_{-35}$ \\
 & 5 & 2.11 & 1296 & (2.5--2.75)$^{+0.5}_{-1.0}$  & (1.5--1.75)$^{+0.25}_{-0.25}$
 	& (110--150)$^{+470}_{-35}$ \\
 & 6 & 2.57 & 1152 & (2.5--2.75)$^{+0.25}_{-1.0}$ & (1.5--1.75)$^{+0.25}_{-0.25}$
 	& (110--150)$^{+470}_{-30}$ \\
 & 7 & 3.0 & 1008 & (2.5--2.75)$^{+0.25}_{-1.0}$ & (1.5--1.75)$^{+0.25}_{-0.25}$
 	& (110--150)$^{+470}_{-30}$ \\
 & 8 & 3.44 & 864 & (2.75--3.0)$^{+0.5}_{-0.5}$  & (1.75--2.0)$^{+0.25}_{-0.25}$
 	& (130--170)$^{+85}_{-40}$ \\
 \hline
 corrected\tablenotemark{c}
 & 1 & 0.33 & 1008 & (3.0--3.25)$^{+0.25}_{-0.5}$  & (2.0--2.25)$^{+0.25}_{-0.25}$
 	& (155--200)$^{+70}_{-40}$ \\
 & 2 & 0.75 & 1296 & (2.75--3.0)$^{+0.25}_{-0.25}$ & (2.0--2.25)$^{+0.25}_{-0.25}$
 	& (160--200)$^{+70}_{-35}$ \\
 & 3 & 1.2 & 1296 & (3.0--3.25)$^{+0.25}_{-0.5}$  & (2.0--2.25)$^{+0.25}_{-0.25}$
 	& (155--200)$^{+70}_{-40}$ \\
 & 4 & 1.67 & 1440 & (3.0--3.25)$^{+0.5}_{-0.5}$   & (1.75--2.0)$^{+0.25}_{-0.25}$
 	& (125--160)$^{+60}_{-45}$ \\
 & 5 & 2.11 & 1296 & (3.0--3.25)$^{+0.5}_{-0.25}$  & (1.75--2.0)$^{+0.25}_{-0.5}$
 	& (125--160)$^{+40}_{-55}$ \\
 & 6 & 2.57 & 1152 & (3.0--3.25)$^{+0.25}_{-0.5}$  & (1.75--2.0)$^{+0.25}_{-0.0}$
 	& (125--160)$^{+60}_{-10}$ \\
 & 7 & 3.0 & 1008 & (3.0--3.25)$^{+0.25}_{-0.25}$ & (2.0--2.25)$^{+0.25}_{-0.25}$
 	& (155--200)$^{+50}_{-40}$ \\
 & 8 & 3.44 & 864 & (3.5--3.75)$^{+0.25}_{-0.5}$  & (2.5--2.75)$^{+0.5}_{-0.5}$
 	& (200--260)$^{+110}_{-90}$ \\
 \enddata
 \tablecomments{The values reported for log$G_{0}$, log($n$/cm$^{-3}$), and T show the best fitting range from the model grid in brackets.  The lower limits on these values are calculated by subtracting the lower uncertainty from the lower end of the best fitting range, while the upper limits should be calculated by adding the upper uncertainty to the upper end of the best fitting range.}
 \tablenotetext{a}{Angular distance from the center of the galaxy to the approximate center of each bin, along an orientation angle of $\sim 116^{\circ}$ east of north.  We remind the reader that the bins are shown schematically in Figure~\ref{fig:schematic}.}
 \tablenotetext{b}{The uncorrected results include all of the observed [C~\textsc{ii}] emission.}
 \tablenotetext{c}{The corrected results include only [C~\textsc{ii}] emission from neutral gas, and the [O~\textsc{i}]63 has been increased by a factor of two to account for multiple PDRs.}
\end{deluxetable}

\subsection{Results}\label{discussion}
In \citet{2012MNRAS.422.2291P} a radially decreasing trend in both the dust temperature and the gas-to-dust mass ratio was reported, implying some influence on the surrounding ISM by the central AGN in Cen~A.  Interestingly, we do not see a radial trend in the density of hydrogen nuclei in PDRs, $n$, nor the strength of the interstellar radiation field impinging onto the PDR surfaces, $G_{0}$.  Even within one standard deviation of the mean value in each bin, there is little trend with increasing radius from the center (only the outermost bin shows a significant deviation from the other bins; however, this may be due to the small number of pixels in that bin).  Correspondingly, the surface temperature of the PDRs also does not show a radial trend, in contrast to the dust temperature, which decreases from 26.5~K in the innermost regions of the area outlined in Figure~\ref{fig:schematic} to 20.5~K in the outermost region.  These results suggest that the physical properties of the molecular clouds nearest the center in our observations are not being affected strongly by the AGN.

The discrepancy between the PDR temperature and the dust temperature may be explained in part by estimating the `observed' value of $G_{0}$ using the total infrared intensity.  Following the method presented by \citet{2005A&A...441..961K}, we calculate the average observed value of $G_{0}$ in each bin across the eastern disk of Cen~A and present the results in Table~\ref{tbl:G0_obs}.  Interestingly, the FUV radiation field predicted by the dust emission is consistent within uncertainties with that determined by the model for all bins except the outermost bin.  This likely indicates that PDRs dominate the total infrared emission in the most, if not all of the eastern disk.  The value of $G_{0}$ calculated using the observed TIR integrated intensity is lower than that determined by the PDR model in the outermost bin.  This could be due to a lower beam filling factor for the dust emission associated with PDRs, compared to the colder, more diffuse dust component.

\begin{deluxetable}{lcc}
\tabletypesize{\small}
\tablecolumns{3}
\tablecaption{Observed and PDR model Predicted Values of $G_{0}$ for the Eastern Disk of Cen~A.\label{tbl:G0_obs}}
\tablewidth{0pt}
\tablehead{
\colhead{Bin} &  \multicolumn{2}{c}{log$G_{0}$} \\
	\colhead{} & \colhead{Average Observed} & \colhead{PDR Model Predicted}}
  \startdata
  1 & $2.3 \pm 2.0$ & (2.0--2.25)$^{+0.25}_{-0.25}$ \\
  2 & $2.3 \pm 1.9$ & (2.0--2.25)$^{+0.25}_{-0.25}$ \\
  3 & $2.3 \pm 2.0$ & (2.0--2.25)$^{+0.25}_{-0.25}$ \\
  4 & $1.7 \pm 1.5$ & (1.75--2.0)$^{+0.25}_{-0.25}$ \\
  5 & $1.4 \pm 0.8$ & (1.75--2.0)$^{+0.25}_{-0.5}$ \\
  6 & $1.3 \pm 1.1$ & (1.75--2.0)$^{+0.25}_{-0.0}$ \\
  7 & $1.2 \pm 0.9$ & (2.0--2.25)$^{+0.25}_{-0.25}$ \\
  8 & $0.7 \pm 0.3$ & (2.5--2.75)$^{+0.5}_{-0.5}$ \\
  \enddata
\end{deluxetable}

One possible explanation for this might be the high inclination of Cen~A with respect to the line of sight.  Cen~A has an inclination of roughly 75$^{\circ}$ \citep{2006ApJ...645.1092Q} so it is nearly edge on.  If we were only diagnosing the characteristics of clouds from the nearest side of the galaxy, we might not observe any effects of the AGN on the surrounding clouds.  However, while we believe the [O~\textsc{i}]$_{63}$ line is optically thick, [O~\textsc{i}]$_{145}$ is not, and it is unlikely the [C~\textsc{ii}], and $F_{\mathrm{TIR}}$ are optically thick as well.  Thus, it is more likely that any effects the AGN might have on the surrounding gas are diluted because we are integrating emission over clouds through the arm and interarm regions as well as any clouds in the vicinity of the AGN along our line of sight.

Another possibility to explain why the dust is affected by the AGN but not the PDR gas is that a significant amount of dust may reside in the diffuse ISM.  This dust could be more strongly affected by the X-rays produced by the AGN than the gas located within PDR regions.

\subsection{Inferred Physical Conditions from PDR Modelling}\label{discuss_pdr}
In Section~\ref{pdr_model} we found that the average values of $G_{0}$ and $n$ across the disk ranged from $\sim 10^{1.75}$--10$^{2.75}$ and $\sim 10^{2.75}$--$10^{3.75}$~cm$^{-3}$, respectively.  These results are consistent with those previously published by \citet{2000A&A...355..885U}, who found $G_{0} \sim 10^{2}$ and $n \sim 10^{3}$~cm$^{-3}$, as well as by \citet{2001A&A...375..566N} who found $G_{0} = 10^{2.7}$ and $n = 10^{3.1}$~cm$^{-3}$, for Cen~A.  The properties of the molecular clouds are also consistent with those found by large surveys on global scales.  The 60 galaxies in the \citet{2001ApJ...561..766M} sample show $10^{2} \le G_{0} \le 10^{4.5}$ and $10^{2}\,\mathrm{cm}^{-3} \le n \le 10^{4.5}\,\mathrm{cm}^{-3}$, while the full sample of \citet{2001A&A...375..566N} shows a range of 10$^{2}$ to 10$^{4}$ for both $n$ and $G_{0}$, where $n$ is in units of cm$^{-3}$.  

\begin{figure}
\includegraphics[width=\columnwidth]{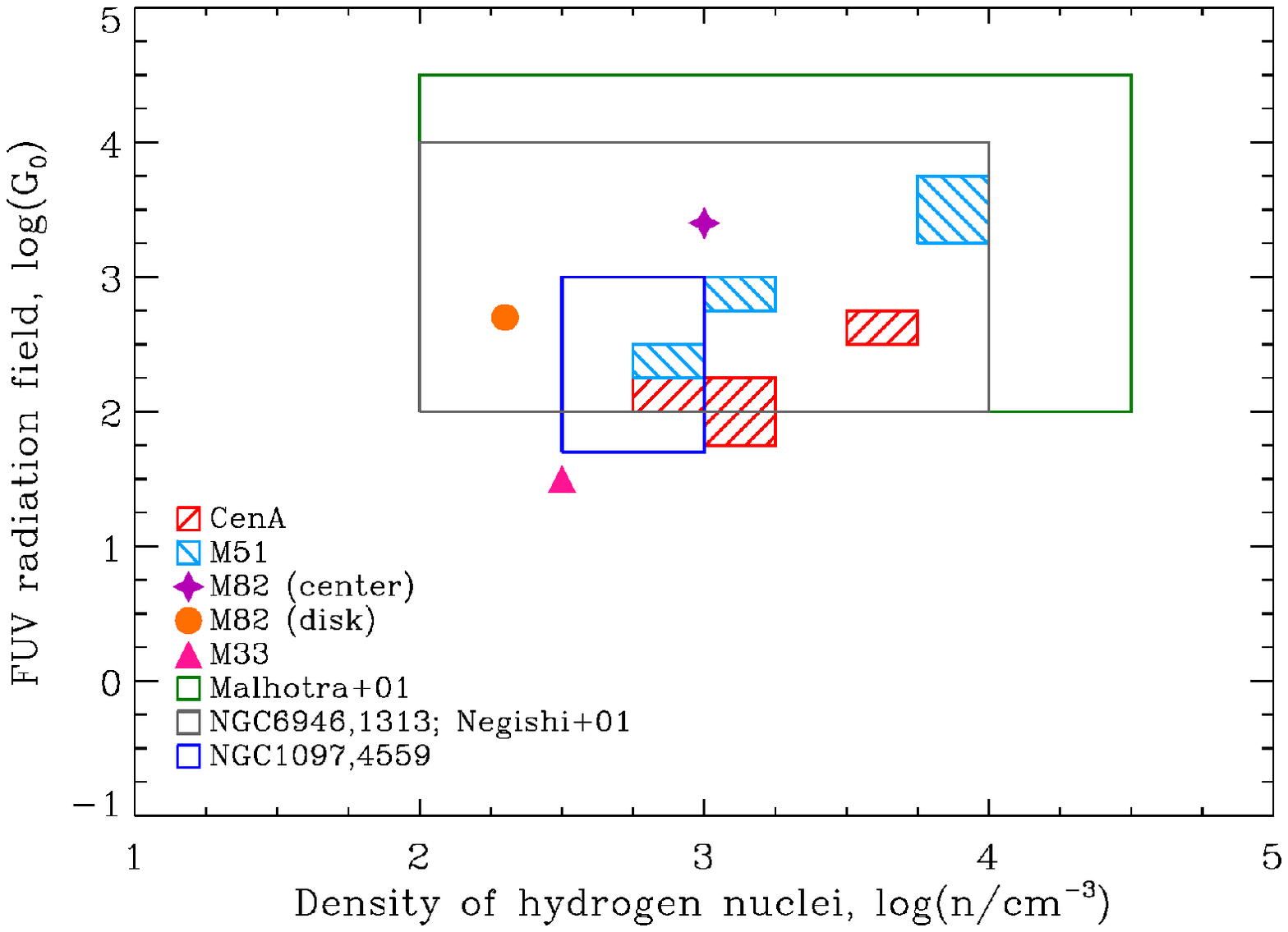}
\caption{The locations of Cen~A and M51 in the PDR characteristic parameter space $G_{0}$-$n$ in comparison to other galaxies.  References: Cen~A (present work), M51 \citep{parkin_2013}, M82 \citep{2013A&A...549A.118C}, NGC1097 and NGC4559 \citep{2012ApJ...747...81C}, NGC~6946 and NGC~1313 \citep{2002AJ....124..751C}, M33 \citep{2011A&A...532A.152M}, and surveys \citet{2001ApJ...561..766M} and \citet{2001A&A...375..566N}.}
\label{fig:G0vsn}
\end{figure}

We also compare our results to those found for other individual galaxies.  In Figure~\ref{fig:G0vsn} we plot the locations of Cen~A, M51, and several other galaxies for which PDR characteristics are available in the literature, in the $G_{0}$-$n$ parameter space.  The PDRs in Cen~A are consistent with those of the Seyfert~1 galaxy NGC~1097 \citep{2012ApJ...747...81C}, and the spiral galaxies NGC~4559 \citep{2012ApJ...747...81C}, NGC~6946 and NGC~1313 \citep{2002AJ....124..751C}, and M33 \citep{2011A&A...532A.152M} (within uncertainties).  We also see that Cen~A has a lower value for $G_{0}$ than the starburst of M82 \citep{2013A&A...549A.118C}, but a higher value for $G_{0}$ than M33.  Overall, Cen~A is in fairly good agreement with the range of values of $G_{0}$ and $n$ found in other sources.

Given that Cen~A is an elliptical galaxy that has merged with a disk galaxy, it is useful to also compare it to other early-type galaxies.  In contrast to the values for $G_{0}$ and $n$ found in normal or starbursting galaxies, \citet{2013ApJ...776L..30W} investigated the elliptical galaxy NGC~4125 and found that the [C~\textsc{ii}]/[O~\textsc{i}]$_{63}$ line ratio is greater than 3--4 (including calibration uncertainties) and ([C~\textsc{ii}]+[O~\textsc{i}]$_{63}$)/$F_{\mathrm{TIR}}$ is greater than $1.3 \times 10^{-3}$.  These values would place it left and upwards in the parameter space in the top right panel of Figure~\ref{fig:pdr_plots}, in a region where only the low $G_{0}$, high $n$ solutions lie (not shown in the figure).  We choose to compare these line ratios for NGC~4125 with our uncorrected results for Cen~A because the [C~\textsc{ii}] emission from NGC~4125 was not corrected for ionized gas.  In fact, it is likely that NGC~4125 is ionized gas dominated given that only an upper limit for [O~\textsc{i}]$_{63}$ is determined but there are significant detections in [N~\textsc{ii}]$_{122}$ and [C~\textsc{ii}] \citep{2013ApJ...776L..30W}.  Furthermore, \citet{2010ApJ...725..100W} find only an upper limit in CO emission for NGC4125.

A study of 12 early type galaxies by \citet{2011MNRAS.410.1197C} and a study focusing on a subsample of early type galaxies from the ATLAS$^{3\mathrm{D}}$ survey \citep{2011MNRAS.414..940Y} by \citet{2012MNRAS.421.1298C}, find that the star forming properties and diagnostic CO line ratios of early type galaxies are similar to normal star-forming, spiral galaxies.  Thus, Cen~A seems to present a``more classical" ISM than NGC~4125 when compared to samples of elliptical and lenticular galaxies, as well as spiral galaxies.


\section{Comparison to M51}\label{compare_m51}
\citet{parkin_2013} investigated the same atomic fine structure lines in the central $\sim2.5\arcmin$ of M51 by dividing the galaxy into four distinct regions: the nucleus, center, arm and interarm regions.  They discovered a radial trend in both the fraction of ionized gas (from about 80\% in the central region of the galaxy down to 50\% in spiral arm and interarm regions) as well as in the properties of the molecular clouds, $n$, $G_{0}$ and $T$.  However, they also discovered that in addition to the radial trend, the molecular clouds in the arm and interarm regions displayed the same physical characteristics, despite differing star formation rate surface densities.  We now discuss the similarities and differences between the properties of the gas in both M51 and Cen~A.

To give any meaning to this comparison we first need to consider the star formation rate (SFR) and star formation rate surface density (SFRD).  We estimate the SFR of Cen~A by using the equation derived empirically by \citet{2013ApJ...768..180L} that uses the luminosity of the \emph{Herschel} PACS 70~$\mu$m map (their equation~(4), with the calibration constant determined for their combined dataset as listed in their Table~5):
\begin{equation}\label{eqn:sfr}
\mathrm{SFR}(70) = 1.18 \times 10^{-43} L(70),
\end{equation}
where the SFR rate is given in M$_{\odot}$~yr$^{-1}$ and the luminosity at 70~$\mu$m is given in erg~s$^{-1}$.  With this equation, we obtain a total SFR in the region outlined in Figure~\ref{fig:schematic} of approximately 0.29~M$_{\odot}$~yr$^{-1}$.   \citet{2013ApJ...768..180L} caution that the SFR determined with this equation may be overestimated by up to 50\%.  Indeed, Equation~\ref{eqn:sfr} assumes that $L(70)$ is entirely associated with recent star formation, while up to half of the observed $L(70)$ could be associated with the older stellar population.  Thus, we conservatively assume half of the 70~$\mu$m is not associated with current star formation.  This leads to an observed SFR to $\sim0.14$~M$_{\odot}$~yr$^{-1}$.

Next we need to estimate the SFRD for Cen~A.  The inclination angle of Cen~A is approximately $i = 75^{\circ}$ \citep{2006ApJ...645.1092Q}, thus we divide the observed area in Figure~\ref{fig:schematic} ($\sim 12.3$~kpc$^{2}$) by $\cos i$ to estimate the physical area of the region covered by the fine structure lines. This gives a SFRD of $\Sigma(70) = 0.01$~M$_{\odot}$~yr$^{-1}$~kpc$^{-2}$.  For consistency, we apply the same equation to M51 using only the area covered by the fine structure line observations in \citet{parkin_2013}, which is roughly 49~kpc$^{2}$.  We obtain a value for the SFR in this region of M51 of 4.85~M$_{\odot}$~yr$^{-1}$, thus a SRFD of $\sim 0.05$~M$_{\odot}$~yr$^{-1}$~kpc$^{-2}$ (again assuming 50\% of the 70~$\mu$m is from recent star formation).   In comparison, \citet{1998ApJ...498..541K} reports a global mean SFR density of approximately 0.02~M$_{\odot}$~yr$^{-1}$~kpc$^{-2}$ for M51, while \citet{2007ApJ...671..333K} find a range of SFRDs between 0.001 and 0.4~M$_{\odot}$~yr$^{-1}$~kpc$^{-2}$ for 257 apertures centred on H~\textsc{ii} regions within M51.  Thus, the the SFRD in Cen~A is lower when compared with the region of M51 mapped in the atomic fine structure lines; however, we caution that there are large uncertainties in these estimates and thus the difference may not actually be as significant.

With the SFRDs of both galaxies in mind, we can first compare the heating efficiency as a function of the 70~$\mu$m/160~$\mu$m ratio between the two galaxies.  In \citet{parkin_2013} it was shown that the average value for the heating efficiency was about $5 \times 10^{-3}$ in the arm and interarm regions of M51, decreasing to $3 \times 10^{-3}$ in the nucleus.  In Figure~\ref{fig:heat_eff} we see that this ratio is slightly higher in Cen~A than in M51, with a value of $5 \times 10^{-3}$ in the bins closest to the AGN, a peak of $7.5 \times 10^{-3}$ in the middle of the strip, and a value of $6 \times 10^{-3}$ in the outermost bins.

\begin{figure}
\includegraphics[width=\columnwidth]{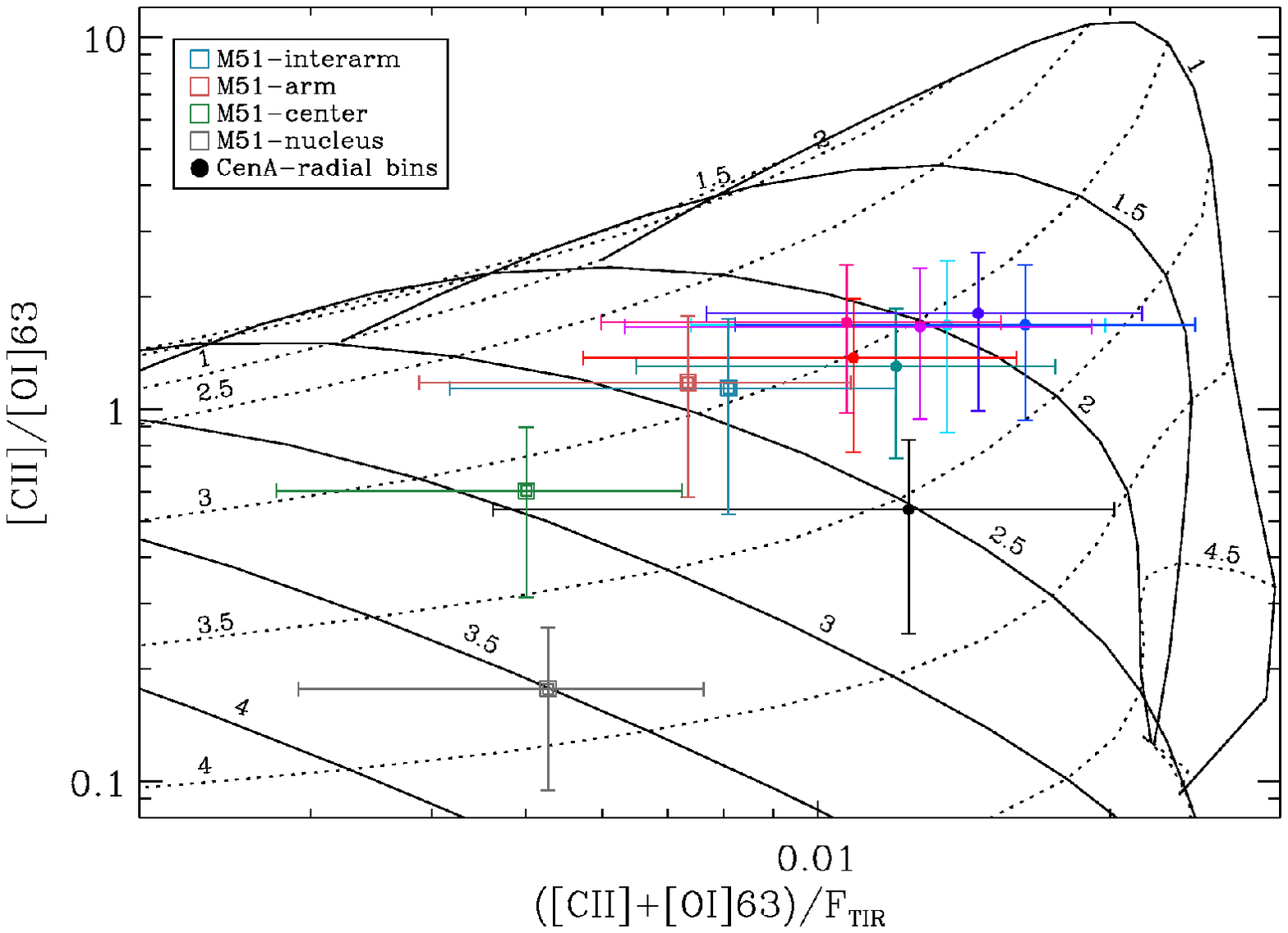}
\caption{A comparison of the PDR model results for the radial bins of Cen~A with the PDR model results for the four regions in the inner region of M51 from \citet{parkin_2013}.}
\label{fig:pdr_comparison}
\end{figure}

Next, we compare the mean values of the [C~\textsc{ii}]/[O~\textsc{i}]$_{63}$ and ([C~\textsc{ii}]+[O~\textsc{i}]$_{63}$)/$F_{\mathrm{TIR}}$ line ratios for each of the four regions of M51 with those from each of the radial bins of Cen~A on the PDR model [C~\textsc{ii}]/[O~\textsc{i}]$_{63}$ versus ([C~\textsc{ii}]+[O~\textsc{i}]$_{63}$)/$F_{\mathrm{TIR}}$ parameter space in Figure~\ref{fig:pdr_comparison}.  The values for $n$ and $G_{0}$ in Cen~A are generally consistent with those of the arm and interarm regions of M51 within uncertainties, even for the innermost bins in our observations.  However, the nuclear and center regions of M51 have slightly higher values for $n$ and $G_{0}$ ($10^{3}$--$10^{4}$~cm$^{-3}$, and $10^{2.75}$--$10^{3.75}$, respectively) and a higher ionized gas fraction (by up to a factor of 4) than observed in the disk of Cen~A (see Figure~\ref{fig:G0vsn}).  This is an interesting result because both galaxies have active centers, with M51 containing a Seyfert~2 nucleus \citep{1997ApJS..112..315H}; thus, we might expect similar properties in their central regions.  We note that we do not have observations directly of the nucleus of Cen~A; however, the result stands even if we ignore the nucleus region of M51 because the center region contains molecular clouds with higher density and is exposed to a stronger radiation field than in Cen~A.  This might also be due to the higher fraction of ionized gas in M51 than in Cen~A.  If M51 has a larger population of massive young stars, they would produce more FUV radiation and thus more H~\textsc{ii} regions.  Alternatively, the difference could be another consequence of the high inclination of Cen~A.  If there is a stronger radiation field affecting clouds near the center of the galaxy, it may be diluted by the weaker fields contributing along the line of sight.  Investigating additional galaxies with active nuclei could confirm which is the more likely scenario.

The [O~\textsc{iii}]/[N~\textsc{ii}]$_{122}$ line ratio indicates that the youngest stars in Cen~A are hotter (O9.5 or O9) than in M51 \citep[B0; ][]{parkin_2013} based on the stellar classifications from \citet{1996ApJ...460..914V}.  However, this apparent discrepancy may be due to lower signal-to-noise in the M51 observations of the [O~\textsc{iii}] line than in Cen~A, suggesting that the observed [O~\textsc{iii}]/[N~\textsc{ii}]$_{122}$ ratio in M51 from \citet{parkin_2013} should be considered as a lower limit.  It is also possible that the ratio observed in Cen~A is a combined effect of the different stellar populations and of a stronger AGN, thus making the apparent stellar classification earlier than it really is.

We conclude that the physical characteristics of the PDRs in the molecular clouds of Cen~A are reasonably similar to those found in normal, star forming galaxies, although there seem to be a few noticeable differences between it and M51 based on our two data sets.

\section{Conclusions}\label{conclusions}
We have presented new spectroscopic observations of the unusual elliptical galaxy Centaurus~A from the \emph{Herschel} PACS instrument.  These observations focus on important atomic cooling lines originating from both neutral ([C~\textsc{ii}](158~$\mu$m), [O~\textsc{i}](63 and 145~$\mu$m)) and ionized gas ([N~\textsc{ii}](122 and 205~$\mu$m) and [O~\textsc{iii}](88~$\mu$m)) covering a radial strip on the eastern side of the nucleus of the galaxy (or a central aperture for the [N~\textsc{ii}](205~$\mu$m) line).  These lines show very similar morphologies, except the two [O~\textsc{i}] lines that peak toward the center of the galaxy.

We find that the heating efficiency in the disk, represented by the ([C~\textsc{ii}]+[O~\textsc{i}]$_{63}$)/$F_{\mathrm{TIR}}$ line ratio, ranges from $4 \times 10^{-3}$ to $8 \times 10^{-3}$, consistent with values determined in galaxies on global scales, as well as on resolved scales in other individual galaxies.  However, we do not observe a significant decrease in the heating efficiency with increasing dust temperature, as represented by the 70~$\mu$m/160~$\mu$m color, nor do we observe a suppression in the heating efficiency in the vicinity of the nucleus.  We also find that the heating efficiency is slightly higher in Cen~A than the grand-design spiral galaxy, M51, suggesting the dust grains and PAHs in Cen~A are more neutral in PDRs than in M51.  Furthermore, the line ratio [O~\textsc{iii}]/[N~\textsc{ii}]$_{122}$ reveals that the youngest stars are of a slightly earlier stellar type than those in M51, thus producing a harder radiation field in the disk of Cen~A. However, there is a possibility that the AGN is partially contributing to the observed emission, resulting in an earlier stellar type classification than is actually present.

A comparison between a PDR model and our observations reveals that the strength of the FUV radiation field incident on the PDR surfaces ranges from $\sim 10^{1.75}$--10$^{2.75}$ and the hydrogen gas density ranges from $\sim 10^{2.75}$--$10^{3.75}$~cm$^{-3}$, in agreement with typical values in other star forming galaxies, including M82, which has a central starburst.  However, the conditions (PDRs) producing the fine structure lines in Cen~A are distinctly different from the elliptical galaxy NGC~4125, where the gas may be completely ionized.  Contrary to M51, we do not see a significant radial trend in either $n$ or $G_{0}$.  Furthermore, while the results from the PDR modelling for Cen~A agree with those for the arm and interarm regions in M51, the central region of M51 shows higher values for $n$ and $G_{0}$.  Observations of the nucleus of Cen~A in the important fine structure lines may reveal a similar trend; however, we point out that in the central region of M51 up to 70\% of the [C~\textsc{ii}] emission originates in diffuse ionized gas while in Cen~A this fraction is only 10--20\%, thus this may also explain the differences between the two galaxies.  We conclude that the disk of Cen~A exhibits properties in its PDRs that are similar to other normal disk galaxies, despite its unusual morphological characteristics.



\acknowledgments
T.~J.~P. thanks the anonymous referee for his/her comments, which have been beneficial to the research presented in this paper.  The research of C.~D.~W. is supported by the Natural Sciences and Engineering Research Council of Canada and the Canadian Space Agency.  I.~D.~L is a postdoctoral researcher of the FWO-Vlaanderen (Belgium).  PACS has been developed by a consortium of institutes led by MPE (Germany) and including UVIE (Austria); KU Leuven, CSL, IMEC (Belgium); CEA, LAM (France); MPIA (Germany); INAF-IFSI/OAA/OAP/OAT, LENS, SISSA (Italy); IAC (Spain). This development has been supported by the funding agencies BMVIT (Austria), ESA-PRODEX (Belgium), CEA/CNES (France), DLR (Germany), ASI/INAF (Italy), and CICYT/MCYT (Spain).  SPIRE has been developed by a consortium of institutes led by Cardiff University (UK) and including Univ. Lethbridge (Canada); NAOC (China); CEA, LAM (France); IFSI, Univ. Padua (Italy); IAC (Spain); Stockholm Observatory (Sweden); Imperial College London, RAL, UCL-MSSL, UKATC, Univ. Sussex (UK); and Caltech, JPL, NHSC, Univ. Colorado (USA). This development has been supported by national funding agencies: CSA (Canada); NAOC (China); CEA, CNES, CNRS (France); ASI (Italy); MCINN (Spain); SNSB (Sweden); STFC and UKSA (UK); and NASA (USA).  HIPE is a joint development by the Herschel Science Ground Segment Consortium, consisting of ESA, the NASA Herschel Science Center, and the HIFI, PACS and SPIRE consortia.  The James Clerk Maxwell Telescope is operated by The Joint Astronomy Centre on behalf of the Science and Technology Facilities Council of the United Kingdom, the Netherlands Organisation for Scientific Research, and the National Research Council of Canada. This work is based, in part, on observations made with the Spitzer Space Telescope, obtained from the NASA/ IPAC Infrared Science Archive, both of which are operated by the Jet Propulsion Laboratory, California Institute of Technology under a contract with the National Aeronautics and Space Administration.  This research has made use of the NASA/IPAC Extragalactic Database (NED) which is operated by the Jet Propulsion Laboratory, California Institute of Technology, under contract with the National Aeronautics and Space Administration.  This research made use of APLpy, an open-source plotting package for Python hosted at http://aplpy.github.com.



{\it Facilities:} \facility{Herschel (PACS, SPIRE)}.

\end{document}